\documentclass[lettersize,journal]{IEEEtran}
\usepackage{amsmath,amsfonts}
\usepackage{algorithm, algpseudocode}
\usepackage{array}
\usepackage[caption=false,font=normalsize,labelfont=sf,textfont=sf]{subfig}
\usepackage{textcomp}
\usepackage{stfloats}
\usepackage{url}
\usepackage{verbatim}
\usepackage{graphicx}
\usepackage{cite}
\usepackage{natbib}
\setcitestyle{numbers,square}
\newtheorem{definition}{Definition}
\newcommand{\ie}{\emph{i.e.}}
\newcommand{\eg}{\emph{e.g.}}
\usepackage{url}
\usepackage{graphicx}
\usepackage{float}
\usepackage{epsfig}
\usepackage{epstopdf}
\usepackage{xspace}
\usepackage{url}
\usepackage{booktabs,etoolbox}
\usepackage{booktabs}
\usepackage{hyperref}
\usepackage{amsfonts}
\usepackage{bm}
\usepackage{booktabs}
\usepackage{tabularx}
\usepackage{array}
\usepackage{rotating}
\usepackage[normalem]{ulem}
\usepackage{nicefrac}
\usepackage{authblk}
\usepackage{microtype}
\usepackage{amsmath, amssymb, amsfonts}
\usepackage{caption}
\usepackage{subcaption}
\usepackage{lipsum}
\usepackage{xcolor}
\usepackage{multirow}
\usepackage{lscape}
\newtheorem{proposition}{Proposition}
\newtheorem{assumption}{Assumption}
\newcolumntype{Y}{>{\raggedright\arraybackslash}X}

\newcommand{\AAA}[1]{{\color{black} #1}}
\begin{document}

\title{NExt-LMM: a two-stage LMM framework with near-exact  variance components for genome-wide association studies}

\author{Zhibin Pu,~\IEEEmembership{Student Member,~IEEE,}
        Shufei Ge,~\IEEEmembership{Member,~IEEE,}
        Shijia Wang,~\IEEEmembership{Member,~IEEE}
\thanks{Manuscript received ; revised. 
}
\thanks{Zhibin Pu, Shufei Ge and Shijia Wang are with the Institute of Mathematical Sciences, ShanghaiTech University, 393 Middle Huaxia Road, 201210, Shanghai, China. 
Correspondence send to Drs. Shijia Wang (wangshj1@shanghaitech.edu.cn) and Shufei Ge (geshf@shanghaitech.edu.cn). 
}
\thanks{Digital Object Identifier:}

}

\markboth{Journal of \LaTeX\ Class Files,~Vol.~14, No.~8, August~2021}%
{Shell \MakeLowercase{\textit{et al.}}: A Sample Article Using IEEEtran.cls for IEEE Journals}


\maketitle

\begin{abstract}
Linear mixed models (LMMs) are widely used in genome-wide association studies (GWAS) to account for population stratification and cryptic relatedness. However, their practical application remains computationally challenging because parameter estimation and association testing require large-scale operations on genetic similarity matrices (GSMs). Here, we present NExt-LMM, a two-stage framework for scalable LMM-based GWAS. In the first stage, NExt-LMM obtains a preliminary estimate of a shared variance ratio under the null model. In the second stage, it constructs the corresponding covariance matrix and accelerates downstream inference through a Hierarchical Off-Diagonal Low-Rank (HODLR) representation. Specifically, NExt-LMM exploits the low-rank structure of off-diagonal covariance blocks to enable substantially faster matrix factorization and inverse-related computations than conventional dense decompositions. It then reuses the resulting HODLR-based covariance representation to accelerate maximum-likelihood-based inference under the shared variance-ratio setting. We further establish error bounds that account for both sources of approximation in the framework: the preliminary variance-ratio estimate and the HODLR approximation. In particular, we show that, conditional on a fixed variance ratio, the NExt-LMM estimator can be made arbitrarily close to the exact shared-variance estimator as the matrix approximation tolerance approaches zero. \AAA{Numerical experiments show that NExt-LMM improves computational efficiency while maintaining high accuracy relative to existing methods, and that the size of this improvement is governed by the numerical rank retained at the tolerance required for calibrated association inference.} We provide an open-source Python implementation at \url{https://github.com/ZhibinPU/NExt-LMM}.
\end{abstract}


\section{Introduction}
\IEEEPARstart{G}{enome}-wide association studies (GWAS) are designed to identify genotype-phenotype associations by examining genetic variants. Classical linear regression methods quantify these relationships but yield false positives due to the population structure. Linear mixed models (LMM) correct the confounders by using measures of genetic similarity as a random effect.  Such measures are often represented by genetic similarity matrices (GSM). However, the GSM inversion in maximum likelihood estimation (MLE) creates computational expense at $\mathcal{O}(n^3)$ complexity. This bottleneck constrains large scale applications where GWAS analyze small-effect variants (\ie~Alzheimer's Disease, AD) using millions of SNPs on large cohorts.

{Several strategies have been developed to scale linear mixed models for GWAS. One class of methods reduces the cost of repeated covariance operations through matrix decomposition or low-rank structure, as exemplified by compressed MLM and FaST-LMM \citep{zhang2010mixed, lippert2011fast}. In particular, FaST-LMM uses a one-time spectral decomposition of the GSM to reparameterize the model and amortize the computational cost across genome-wide marker testing. Another widely used class adopts a two-stage approximation strategy, in which variance components are estimated once under a null model and then reused during marker-wise association testing. Representative examples include P3D \citep{zhang2010mixed}, BOLT-LMM \citep{loh2015efficient} , and fastGWA \citep{jiang2019resource}. Among these, fastGWA achieves scalability by constructing and exploiting a sparse GSM, typically by retaining only sufficiently large pairwise genetic relationships, and then leveraging sparse-matrix operations for efficient mixed-model association testing. This sparse-matrix strategy is fundamentally different from low-rank approximations \citep{mbatchou2021computationally}.}

{These methods have greatly improved the practical scalability of LMM-based GWAS, but they rely on different computational tradeoffs. Spectral methods such as FaST-LMM incur a substantial one-time decomposition cost, although the subsequent per-marker scan is efficient. Sparse-GSM methods such as fastGWA depend on the degree of sparsity in the realized relatedness structure. Iterative approaches such as BOLT-LMM avoid explicit dense matrix factorization, but their cost remains nontrivial at large sample size. These considerations motivate the search for alternative matrix representations that can accelerate covariance operations while maintaining high numerical fidelity.}

Although these works have been widely successful in significantly reducing the high computational cost of LMMs in  large scale datasets by optimized implementations and approximation approaches, several concerns still exist about the approximation cost of the GSM in both the computational time and the accuracy. Precisely, for common methods to approximate the GSM, such as spectral decomposition and singular value decomposition \citep{lippert2011fast, pirinen2013efficient, zhou2022generalized}, the complexity is often cube of the cohort size \citep{trefethen2022numerical}, which is quite expensive for large-scale datasets. Moreover, there is a lack of theoretical analysis of the error between the estimators of the approximated likelihood and the original one, making it hard to provide insights on the accuracy of the heuristics that have been used to speed up LMM computations.

Recently, a hierarchical off-diagonal low-rank (HODLR) approximation was developed and applied to handle the computational bottlenecks in matrix inversion problems \citep{ambikasaran2015fast, williams2020precessing, moran2022fast}. In particular, \AAA{when the off-diagonal blocks admit low-rank approximations at the prescribed tolerance,} the HODLR method could largely decrease the computational cost from $\mathcal{O}(n^3)$ to almost linear work, with a solid theoretical basis for the error control (where $n$ is the size of the matrix). We also notice that many scalable LMM-based GWAS methods adopt a two-stage strategy: variance components are estimated once under a null or population-level model and then reused across markers. Therefore, we follow this practical regime and focuses on a different bottleneck: accelerating the covariance operations within that shared-variance LMM scan. In this article, we introduce the HODLR method into the two stage LMM framework, and propose a linear mixed model with near-exact covariance operations (NExt-LMM) lessening the potential weaknesses of recent LMM works. 

This new framework combines the main properties of the two LMM lines for efficient computations. \AAA{On the one hand, NExt-LMM represents the GSM-induced covariance matrix using a HODLR approximation with tolerance-dependent off-diagonal ranks. For a fully recursive HODLR implementation, and when these ranks remain bounded, standard HODLR theory yields an inversion cost of $\mathcal{O}(k_\epsilon^2 n \log^2 n)$, rather than $\mathcal{O}(n^3)$, where $k_\epsilon$ denotes the rank retained at tolerance $\epsilon$. The practical gain, however, depends on the numerical compressibility of the realized covariance matrix and on the tolerance required for calibrated inference. We therefore evaluate this accuracy--storage--runtime trade-off empirically in Section~\ref{sec:sim}.} \AAA{We also investigate the theoretical error of the NExt-LMM estimator, and establish perturbation bounds showing that its marker-effect and residual-variance estimators converge to those of the exact shared-$\lambda$ two-stage estimator at a rate linear in the compression tolerance $\epsilon$.} On the other hand, this method takes the insight that the variance parameters could be assumed the same across all the SNPs, and applies a shared ratio of the heritable and non-heritable variance parameters to speed up the likelihood estimation.

The rest of this article is organized as follows. In Section \ref{sec:med}, we introduce our proposed NExt-LMM, and show some theoretical properties of the approach. Section \ref{sec:sim} uses numerical results to show the effectiveness of
our method. In Section \ref{sec:con}, we summarize our work.

\section{Methods}
\label{sec:med}

\subsection{Linear mixed model}

The linear mixed model (LMM) aims to identify the association between genetic markers and the phenotype.
Independent LMMs may be applied at each SNP (single nucleotide
polymorphisms) as follows:
{
\begin{align}
    \textbf{y} = W_p\bm{\theta}_p + \textbf{g}_p +\bm{\epsilon}_p,
\end{align}
where $\textbf{y} \in \mathbb{R}^n$ is a column vector of phenotype for $n$ samples. $W_p = [X_C, x_p]$ is the design matrix with $X_C \in \mathbb{R}^{n\times c}$ denoting the covariate matrix and $x_p \in \mathbb{R}^n$ denoting genotype data observed at the $p$-th SNP for all $n$ subjects. $\bm{\theta}_p = (\alpha^T, \beta_p)^T$ is the effect vector, where $\alpha$ corresponds to the covariate effects and $\beta_p$ is the effect estimate for the tested SNP $x_p$. $\textbf{g}_p$ denotes the random effects of the $p$-th SNP and $\bm{\epsilon}_p$ is the random error vector.} The random effects vector $\textbf{g}_p\sim  \text{MVN}(\bm{0}, K\cdot\sigma_g^2)$ and 
the random error $\bm{\epsilon}_p \sim  \text{MVN}(\bm{0}, I_n\cdot\sigma_e^2)$ are assumed as independent vectors with the covariance scalars $\sigma^2_g$ and $\sigma^2_e$ as the heritable and non-heritable components, respectively. 
The genetic similarity matrix (GSM) 
$K$ measures the genetic relatedness among the $n$ samples. It is an $n \times n$ positive semi-definite matrix constructed as {$K = XX^\top/ P$, where $X = (x_1, \ldots, x_P) \in \mathbb{R}^{n \times P}$ denotes the whole column-standardized genotype matrix, and $P$ is the number of SNPs.} 

To improve efficiency, most LMM frameworks do not directly choose $\sigma^2_g$ and $\sigma^2_e$ as the optimization parameters. Instead, the unknown effects with the heritability $h^2 = \frac{\sigma_g^2}{\sigma_e^2 + \sigma_g^2}$ or the ratio $\lambda = \frac{\sigma_g^2}{\sigma_e^2}$ are often regarded as parameters to be estimated by maximum likelihood methods. The primary computational limitation in LMMs stems from the dense and unstructured nature of $K$. Precisely, the fundamental operations required for the likelihood maximization, such as the inversion of the covariance matrix $\Sigma = \lambda K+I_n$, exhibit $\mathcal{O}(n^3)$ time complexity without necessary matrix compression methods. For large scale genetic studies where $n > 10^5$, the maximum likelihood estimation would become computationally prohibitive to obtain results on the whole SNP data.

\subsection{Hierarchical off-diagonal low-rank matrices}

With the increasing demands of computational resources for matrix operations in genetic studies, matrix compression for the LMMs has proven to be valuable and ubiquitous. In particular, the GSMs approximately exhibit low-rank structure due to their inherent redundancy or patterns, allowing them to be approximated by compressing the sub-matrices. Several classical methods, including the $Nystr\ddot{o}m$ approximation \citep{geoga2020scalable}, fast multipole method \citep{li2021rapid} and mosaic approximation \citep{setukha2022application}, have been applied to other fields. The resulting approximation is the so-called hierarchical matrices. These matrices are formed by a set of matrix divisions derived from the original one. The essential operations for these matrices (matrix-vector and matrix--matrix multiplication, addition and inversion) can be performed in, up to logarithmic factors, optimal complexity. 

\begin{figure}
    \centering
    \includegraphics[width=1.0\linewidth]{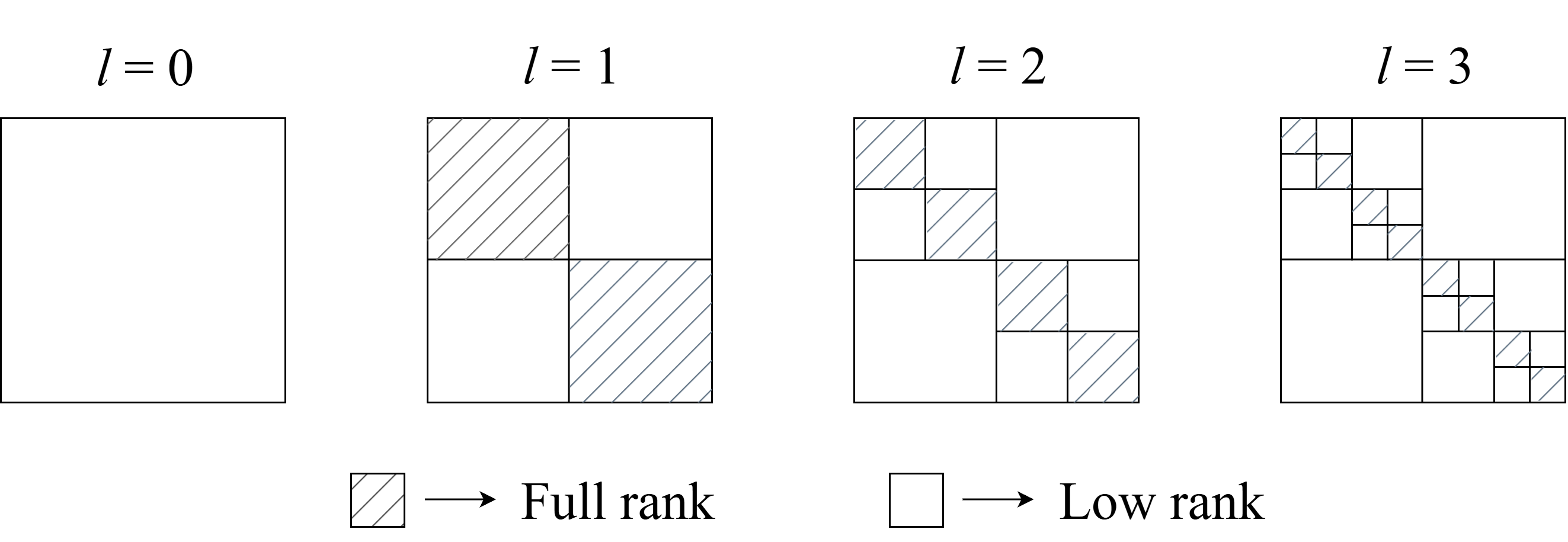}
    \caption{The HODLR matrix in the same size but at different levels.}
    \label{fig: hodlr script}
\end{figure}

In this article, we focus on the class of hierarchical matrices known as Hierarchical Off-Diagonal Low-Rank (HODLR) matrices. As the name
suggests, this class of matrices has off-diagonal blocks that are efficiently represented in a recursive fashion. A graphical representation of this class of matrices is displayed in Figure \ref{fig: hodlr script}. Each panel represents the same matrix, but viewed on different hierarchical scales to show the particular rank structure. 

The HODLR format is particularly suited for square GSM-based covariance matrices. For an arbitrary square matrix $M \in \mathbb{R}^{n \times n}$, we use $\mathcal{H}_{l}(M)$ to denote 
the corresponding HODLR format as a block operator of $M$ at division level $l$, which 
is defined through recursive block partitioning where the minimum sub-block dimension $n/2^{l}$ is determined by $l$. To describe the recursive partitioning scheme, we start from the initial division $ \mathcal{H}_{0}(M) = M$, indicating that the $0$-level block is exactly the full matrix $M$ and no matrix division is done. When $l=1$, the dimension of the minimum block becomes $n/2$, we have the following block partition $\mathcal{H}_{1}(M)$:
\begin{equation}\label{equ1}
\begin{aligned}
    \mathcal{H}_{1}(M) = \begin{bmatrix}
       \mathcal{H}_{0}(M_{11}) & \mathcal{R}_{0}^{(k)}(M_{12}) \\
       \mathcal{R}_{0}^{(k)}(M_{21}) & \mathcal{H}_{0}(M_{22})
    \end{bmatrix}
    = \begin{bmatrix}
       M_{11} & U_{12}V_{12}^T \\
       U_{21}V_{21}^T & M_{22}
    \end{bmatrix},
\end{aligned}
\end{equation}
where $M_{ij} \in \mathbb{R}^{n/2, n/2}$ ($i, j = 1, 2$), $\mathcal{R}_{0}^{(k)}$ is the operator for low-rank approximation of $0$-level blocks on the off diagonal, which can be expressed as the product of $U, V  \in \mathbb{R}^{n/2 \times k}$ based on the $k$-rank ($k \ll n$) approximating technique such as the singular value decomposition (SVD). 

Since the blocks on the diagonal are still real symmetric matrices, a further low-rank approximation could be performed iteratively on $M_{11}$ and $M_{22}$. Suppose that the HODLR operator of $M_{ii} \in \mathbb{R}^{n/2, n/2}$ ($i = 1, 2$) is $\mathcal{H}_{l-1}$, with $\mathcal{R}_{l-1}^{(k)}$ representing the $k$-rank approximation of the $(l-1)$-level off-diagonal blocks, the recursive structure of the format $\mathcal{H}_l$ can be defined by:
\begin{equation}
    \mathcal{H}_{l}(M) = \begin{bmatrix}
       \mathcal{H}_{l-1}(M_{11})  & \mathcal{R}_{l-1}^{(k)}(M_{12})  \\
       \mathcal{R}_{l-1}^{(k)}(M_{21}) & \mathcal{H}_{l-1}(M_{22}) 
    \end{bmatrix}.
\end{equation}
Since the result contains approximation errors, the local rank $k$ should be chosen according to some desired accuracy of the HODLR format. For GSMs, we recommend adaptive rank determination at level $l$, which is detailed in Algorithm $1$. \AAA{Besides, in the hybrid implementation used in the numerical experiments, recursion stops when the block size reaches a prescribed minimum size $s$, and the resulting terminal blocks are retained and inverted densely. We use $s=64$ as the default. We additionally provide a fully recursive implementation, for which the recursion continues to the smallest admissible blocks, so that the block level reaches $l_{max} = \lfloor \log_2 n \rfloor$, and which is designed to attain the standard HODLR complexity under bounded off-diagonal ranks.}

\AAA{Provided that the off-diagonal ranks retained at the prescribed tolerance remain bounded,} the HODLR format enables most matrix operations to be performed with almost linear computational complexity, instead of $\mathcal{O}(n^3)$ or $\mathcal{O}(n^2)$ based on the full matrix \citep{kressner2018fast}. In particular, the complexity of matrix inversion could be decreased to $\mathcal{O}(k^2n \log^2n)$ based on the following block structure for the direct inversion \citep{hackbusch2015hierarchical}, where \(k\) denotes the maximum numerical off-diagonal rank required by the HODLR compression at the prescribed tolerance (The more precise version would be $\mathcal{O}(k_{n,\epsilon}^2 n \log^2 n)$ with the prescribed tolerance $\epsilon$):
\begin{equation}
    M^{-1} = \begin{bmatrix}
        M_{11}^{-1} + M_{11}^{-1}M_{12}S^{-1}M_{21}M_{11}^{-1} &  -M_{11}^{-1}M_{12}S^{-1} \\
        -S^{-1}M_{21}M_{11}^{-1} & S^{-1}
    \end{bmatrix}
\end{equation}
where $S = M_{22}-M_{21}M_{11}^{-1}M_{12} \in \mathbb{R}^{n/2 \times n/2}$ is the Schur complement. 
The algorithmic description of HODLR-inverse is shown in Algorithm \ref{alg: hinverse}. One should notice that the algorithm only works when the diagonal block is regular at any level of partition. However, this assumption is always satisfied for the covariance matrix $\Sigma = \lambda K + I_n$ due to the positive definite property. 

\begin{algorithm}[H]
\caption{HODLR-inverse}
\label{alg: hinverse}
\begin{algorithmic}[1]
  \State \textbf{Input:} Real symmetric matrix $M \in \mathbb{R}^{n \times n}$, the prescribed tolerance $\epsilon$, the size of desired smallest block $s$.
  \State \textbf{Output:} Approximated inversion of $M$.
  \State \textbf{Initialize:} Set the rank of off-diagonal approximation $k=1$.
  \If{size($M$) $<$ $s$}
    \State  Return direct-inv($M$);
  \Else
  \State  Partition $M$ into $M_{11}$, $M_{12}$, $M_{21}$ and  $M_{22}$ as Equation $\ref{equ1}$; 
    
  \State  $M_{11}^{-1} \leftarrow$ Perform HODLR-inverse($M_{11}$, $\epsilon$, $s$); 
  \State  Compute $S \leftarrow$ $M_{22}-M_{21}M_{11}^{-1}M_{12}$;
  \State $S^{-1} \leftarrow$ Perform HODLR-inverse($S$, $\epsilon$, $s$); 
  \State $U$, $V \leftarrow$  Low-rank($M_{12}$, $k$);
  \While{ $\Vert$ $UV^T- M_{12}$ $\Vert_2$ $\geq$ $\epsilon$}  
    \State Set $k \leftarrow k+1$;
    \State Reperform $U$, $V \leftarrow$  Low-rank($M_{12}$, $k$);
  \EndWhile
  \State Compute $Z_{12} \leftarrow -M_{11}^{-1}UV^TS^{-1}$;
  \State Compute $Z_{21} \leftarrow -S^{-1}VU^TM_{11}^{-1}$;
  \State Compute $Z_{11} \leftarrow M_{11}^{-1} + M_{11}^{-1}UV^TS^{-1}VU^TM_{11}^{-1}$;
  \State Return $\tilde{M}^{-1} \leftarrow \begin{bmatrix}
      Z_{11} & Z_{12}\\
      Z_{21} & S^{-1}
  \end{bmatrix}$ 
   \EndIf
\end{algorithmic}
\end{algorithm}

\subsection{A two-stage linear mixed model with near-exact covariance operations}
One potential bottleneck of computing the maximum likelihood (or the restricted maximum likelihood, REML) lies in the difficulty to handle the inversion and determinant of the dense matrix $\Sigma = \lambda K + I_n$, where the ratio $\lambda$ is unknown and is hard to be partitioned from $\Sigma$. Lippert \cite{lippert2011fast} proposed the FaST-LMM by rewriting the maximum likelihood as a function of the single ratio parameter $\lambda$ and perform the spectral decomposition on the GSM. 
Besides, to speed-up a full GWAS scan, FaST-LMM considers the population-level heritability by the maximum likelihood setting for the null model, and reuses this estimated heritability for all alternative models. However, 
the computational cost of GSM decomposition at $\mathcal{O}(n^3)$ level still causes computing burdens for large scale datasets. 

We build our NExt-LMM on the similar insight that the same SNP-based heritability is used across all the SNPs, but propose a more efficient framework jointly with HODLR format to avoid the inefficiency arising from direct decomposition methods on the dense covariance matrix. Our algorithm is composed of two approximation stages: (1) We obtain the shared SNP heritability ratio $\tilde{\lambda}$ using the phenotype correlation--genotype correlation regression (PCGC, \citep{golan2014measuring}). (2) We perform a HODLR approximation of the resulting covariance matrix $\Sigma = \tilde{\lambda}K + I_n$. Then we estimate the fixed effects $\beta$ and the non-heritable variance components $\sigma_e^2$ by fixing $\lambda=\tilde{\lambda}$ at the estimated value.

We now turn to the detailed description of our proposed approach. 
Firstly, we employ the PCGC method for the SNP heritability estimation, which is a computationally efficient alternative to the REML. This choice is motivated by the analytical tractability of the PCGC, which reduces the variance component estimation to solving linear equations. The method leverages the fundamental relationship $E(y_iy_j) = f(h^2, K_{ij})$, where $y_i$, $y_j$ are the $i$-th and $j$-th element of the standardized phenotype vector, with the product representing the phenotypic correlation. $K_{ij}$ is the element in the $i$-th row and $j$-th column of the GSM. $f(\cdot)$ is a user-specific mapping function.  For continuous phenotypes under additive genetic models, this function simplifies to $f(h^2, K_{ij}) = h^2 K_{ij}$. Thus, the heritability estimate $\tilde{h}^2$ is obtained by minimizing the squared discrepancy between observed and predicted phenotypic similarities:
\begin{equation}
\begin{aligned}
    \tilde{h}^2 &= \mathop{\arg\min}_{0 \leq h^2 \leq 1} \sum_{i \neq j} [y_iy_j - h^2 K_{ij}]^2. \\
\end{aligned}
\end{equation}
This quadratic optimization yields a closed-form solution via the least square estimation. Crucially, PCGC only uses off-diagonal GSM elements (interaction terms), reducing the computational complexity to $\mathcal{O}(n^2)$.  Note that although PCGC and REML originate from distinct statistical frameworks (method-of-moments and likelihood-based, respectively), both analytical \citep{chen2016reconciliation}  and empirical \citep{wu2018scalable} studies confirm their consistent estimation of $h^2$ under polygenic architectures with neutral genetic effects. This equivalence validates our use of PCGC for the heritability estimation while maintaining theoretical alignment with our following properties derived based on the likelihood foundations.

Secondly, we estimate $\bm{\beta}$ and $\sigma^2_e$ by sharing
$\tilde{\lambda} = \frac{\tilde{h}^2}{1-\tilde{h}^2}$  across all the SNPs. A biological rationale for using a shared SNP-based heritability is the pervasive polygenic architecture of complex traits. Under the classical infinitesimal model and the more recent omnigenic perspective \citep{yang2010common, barton2016infinitesimal, boyle2017expanded}, thousands of variants of very small effect are distributed genome-wide, so treating SNP effects as exchangeable with a shared variance is a reasonable baseline when functional annotations are not explicitly modelled. This assumption underpins widely used linear mixed models (\ie, GREML \cite{yang2016gcta}, BOLT-LMM \cite{loh2015efficient} and LD Score regression \cite{bulik2015ld}), which have been validated across diverse traits.

The original log likelihood $LL(\bm{\theta}, \sigma_g^2, \sigma_e^2) = \log N(W_p\bm{\theta},  \sigma_g^2 K + \sigma_e^2 I_n)$ is parameterized by the effect size $\bm{\beta}$, the heritable and non-heritable variance components $\sigma_g^2$ and  $\sigma_e^2$. Since we introduce $\lambda =  \sigma_g^2 / \sigma_e^2$ and pre-estimate it with the  PCGC regression, the log likelihood is only parameterized by $\bm{\theta}$ and $\sigma_e^2$ as:
\begin{equation}
\begin{aligned}
    &LL(\bm{\theta}, \sigma_e^2) = \log N(W_p\bm{\theta},  \sigma_e^2 \tilde{\Sigma}) \\
    &= C - \frac{n}{2} \log( \sigma_e^2) - \frac{1}{2\sigma^2_e}
    (\textbf{y} - W_p\bm{\theta})^T\tilde{\Sigma}^{-1}(\textbf{y} - W_p\bm{\theta}),
    \label{eqt: 8}
\end{aligned}
\end{equation}
where $\tilde{\Sigma} = \tilde{\lambda} K + I_n$ is the estimated covariance matrix. We perform the maximum likelihood by setting the gradient of the log likelihood in Eq. (\ref{eqt: 8}) with respect to $\bm{\theta}$ to $\bm{0}$, yielding:
\begin{equation}
\begin{aligned}
    \frac{\partial LL(\bm{\theta}, \sigma_e^2)}{\partial \bm{\theta}}  &= \frac{1}{\sigma_e^2}(\textbf{y}-W_p\bm{\theta})^T\tilde{\Sigma}^{-1}W_p = 0,\\
   \tilde{\bm{\theta}} &= (W_p^T \tilde{\Sigma}^{-1} W_p)^{-1}W_p^T \tilde{\Sigma}^{-1} \textbf{y}. \label{eq:theta}
\end{aligned}
\end{equation}
In this step, since the inverse operation of $\tilde{\Sigma}$ is inevitable, we compute the covariance inverse using the HODLR procedure (Algorithm \ref{alg: hinverse}). \AAA{For a fully recursive implementation with bounded retained off-diagonal ranks, the standard HODLR inversion complexity is $\mathcal{O}(k_\epsilon^2 n \log^2 n)$. The hybrid implementation used in our experiments retains a dense terminal block; its runtime and storage are therefore evaluated empirically rather than interpreted as attaining this asymptotic bound.} Notice that $\tilde{\Sigma}^{-1}$ remains the HODLR format. \AAA{For the fully recursive representation, applying $\tilde{\Sigma}^{-1}$ to a vector has complexity $\mathcal{O}(k_\epsilon n \log n)$ under bounded retained ranks \citep{kressner2018fast}, so that this equation can be evaluated at the same cost if $\tilde{\Sigma}^{-1}$ is pre-calculated. The corresponding cost of the hybrid implementation is evaluated empirically.} 
The REML estimator of $\sigma^2_e$ would be unbiased and more consistent with the estimated heritability by PCGC. Thus, we obtain $\tilde{\sigma}^2_e$ by substituting $\tilde{\bm{\theta}}$ into the restricted log-likelihood function and setting the gradient to $0$:
\begin{equation}
\begin{aligned}
    \frac{\partial LL(\sigma_e^2)}{\partial \sigma_e^2}  &= \frac{\partial}{\partial \sigma_e^2} \{-\frac{1}{2} \log  |\sigma^2_e \tilde{\Sigma}| -\frac{1}{2} \log |\sigma^2_e W_p^T\tilde{\Sigma}W_p| \\
    &-\frac{1}{2} (\textbf{y} - W_p\tilde{\bm{\theta}})^T\tilde{\Sigma}^{-1}(\textbf{y} - W_p\tilde{\bm{\theta}}) \}= 0,
     \\
   \tilde{\sigma}_e^2 &= \frac{1}{n-c-1} (\textbf{y} - W_p\tilde{\bm{\theta}})^T\tilde{\Sigma}^{-1}(\textbf{y} - W_p\tilde{\bm{\theta}}).
   \label{eq:sigma}
\end{aligned}
\end{equation}
 \AAA{Under the same fully recursive convention, this equation can likewise be evaluated in at most $\mathcal{O}(k_\epsilon n \log n)$ if $\tilde{\Sigma}^{-1}$ is pre-calculated.} Algorithm \ref{alg: nelmm} provides an overview of our NExt-LMM approach.  

{We summarize the dominant computational costs of several representative LMM approaches in Table \ref{tab:complexity}. We divide the computation into a one-time setup stage and a genome-wide marker-scanning stage, as these costs are amortized differently across methods. For example, FaST-LMM incurs a substantial one-time spectral decomposition cost, after which marker-wise testing is efficient. Likewise, fastGWA derives its scalability by constructing and exploiting a sparse GSM, so its performance depends on the realized sparsity pattern rather than on low-rank structure. BOLT-LMM, in contrast, avoids explicit dense matrix decomposition and relies on iterative whole-genome regression.}

{Under the same setup-versus-scan perspective, NExt-LMM first obtains a shared variance-ratio estimate and then performs a one-time HODLR factorization of the covariance matrix \(\Sigma(\tilde{\lambda}) = I_n + \tilde{\lambda}K\). This factorization is subsequently reused across marker tests, yielding an approximately \(\mathcal{O}(k_\epsilon n \log n)\) cost per marker once the covariance structure has been constructed. \AAA{For NExt-LMM, the displayed HODLR costs describe the standard fully recursive regime with bounded retained ranks. The hybrid implementation used in the numerical experiments retains dense terminal blocks; its runtime and storage are reported empirically in Section~\ref{sec:sens}.} We emphasize that these methods achieve scalability through different computational mechanisms---spectral decomposition, sparse-matrix algebra, iterative whole-genome regression, or hierarchical low-rank matrix representation and Table \ref{tab:complexity} is intended to compare these dominant computational regimes rather than to suggest that practical GWAS software performs a dense \(\mathcal{O}(n^3)\) inversion for every SNP.}

\begin{table*}[!t]
\centering
\caption{{Dominant computational costs of representative LMM-based GWAS methods. We report the one-time setup cost separately from the genome-wide scan cost to distinguish preprocessing/factorization from per-marker testing. Here, \(n\) is the sample size, \(P\) is the number of markers tested, \AAA{\(k_\epsilon\) is the off-diagonal block rank retained by the HODLR representation at tolerance \(\epsilon\), and \(n_{\mathrm{nz}}\)} denotes the number of nonzero entries in a sparse GSM. Costs are shown at the order-of-magnitude level and may vary somewhat across implementations. \AAA{For NExt-LMM, the displayed HODLR costs describe the standard fully recursive regime with bounded retained ranks. The hybrid implementation used in the numerical experiments retains dense terminal blocks; its runtime and storage are reported empirically in Section~\ref{sec:sens}.}} }
\label{tab:complexity}
\footnotesize
\renewcommand{\arraystretch}{1.15}
\begin{tabularx}{\textwidth}{>{\bfseries}l Y Y Y}
\toprule
Method & One-time setup cost & Genome-wide scan cost & Main computational strategy \\
\midrule
FaST-LMM 
& \(O(n^3)\) spectral decomposition of the GSM 
& \(O(nP)\) 
& Dense spectral decomposition amortized across markers \\

BOLT-LMM 
& Iterative null-model fitting and calibration 
& Approximately linear in \(P\); commonly summarized as superlinear in \(n\) overall 
& Iterative whole-genome regression without explicit dense decomposition \\

fastGWA 
& Construction and factorization of a sparse GSM 
& \(O\!\left((n_{\mathrm{nz}}+n)P\right)\) in the sparse-GSM regime 
& Sparse-matrix operations based on a constructed sparse GSM \\

NExt-LMM 
& Preliminary estimation of \(\tilde{\lambda}\) + HODLR factorization of \(\Sigma(\tilde{\lambda})\); \(O(n^2) + O(k_\epsilon^2 n \log^2 n)\)
& \(O(k_\epsilon n \log n\, P)\)
& Two-stage framework with HODLR covariance representation \\
\bottomrule
\end{tabularx}
\end{table*}

{As summarized in Table \ref{tab:complexity}, the main distinction among these methods lies not only in asymptotic cost but also in the source of computational savings. FaST-LMM amortizes a dense eigendecomposition across all markers, fastGWA exploits sparsity in a constructed sparse GSM, and BOLT-LMM relies on iterative updates that avoid explicit dense matrix factorization. By contrast, NExt-LMM targets the covariance operations in a two-stage shared-variance LMM framework and accelerates them through a HODLR representation of \(\Sigma(\tilde{\lambda})\). This comparison clarifies that our contribution is a different numerical strategy for scalable LMM-based GWAS, rather than an alternative to an impractical per-SNP dense inversion scheme.}

\begin{algorithm}
\caption{NExt-LMM}
\label{alg: nelmm}
\begin{algorithmic}[1]
  \State \textbf{Input:} Normalized design matrix $W_p \in \mathbb{R}^{n \times (c+1)}$, response vector $\textbf{y} \in \mathbb{R}^{n}$, genetic similarity matrix $K \in \mathbb{R}^{n \times n}$, the desired error $\epsilon$, the size of desired smallest block $s$.
  \State \textbf{Output:} Estimated $\tilde{\beta}_p$ and $\tilde{\sigma}_e^2$.
  \State \textbf{Initialize:} Set the rank of off-diagonal approximation $k=1$;
  \State Estimate the population-level ratio of variance components $\tilde{\lambda}$ using PCGC regression; 
   \State Let $\tilde{\Sigma} \leftarrow \tilde{\lambda} K + I_n$.
   \State Compute $\tilde{\Sigma}^{-1} \leftarrow $ HODLR-inverse($\tilde{\Sigma}$, $\epsilon$, s);
   \For {each column $X_p$ in $X$}:
   \State Compute $\tilde{\bm{\theta}} \leftarrow (W_p^T\tilde{\Sigma}^{-1} W_p)^{-1}W_p^T\tilde{\Sigma}^{-1} Y$;
   \State Keep $\beta_p$ in $\tilde{\bm{\theta}} = (\tilde{\bm{\alpha}}, \beta_p)^T$;
   \State Compute $\tilde{\sigma}_e^2 \leftarrow \frac{1}{n-c-1} (\textbf{y} - W_p\tilde{\bm{\theta}})^T\tilde{\Sigma}^{-1}(\textbf{y} - W_p\tilde{\bm{\theta}})$.
   \EndFor
\end{algorithmic}
\end{algorithm}

\subsection{Properties}
\label{sec:prop}

{
In this section, we show several approximation properties of NExt-LMM under mild regularity conditions. We distinguish between the \emph{original LMM}, where the variance parameters are re-estimated separately for each marker during association testing, and the \emph{two-stage LMM}, in which a common variance ratio is estimated once under the null model and then reused across marker-wise association testing. In particular, we name the \emph{two-stage LMM} without approximation on $\Sigma$ as the \emph{exact two-stage LMM}. The NExt-LMM is a special case of \emph{two-stage LMM}s which applies the HODLR structure to approximate $\Sigma$.
The theoretical results presented in this section target the estimators of the exact two-stage LMM.

Here we use the notations from  Section~II C. We let $\lambda_0$ denote the underlying true variance ratio under the exact null-model LMM, and denote the true underlying covariance matrix by $\Sigma(\lambda_0)$. 
We let $\tilde\lambda$ denote a variance ratio  estimate under the null model, let $\Sigma(\tilde\lambda) := I_n+\tilde\lambda K$  denote the covariance matrix of the two-stage procedure with an estimated $\tilde\lambda$, and let $\tilde{\Sigma}(\tilde\lambda)$ denote its HODLR approximation. 

\begin{assumption}
\label{Ass: ass1}
For the covariance matrix $\Sigma(\tilde{\lambda})=I_n+\tilde{\lambda}K$,
there exists a HODLR approximation $\tilde{\Sigma}(\tilde{\lambda})\in\mathcal{H}$ such that
\[
\|\Sigma(\tilde{\lambda})-\tilde{\Sigma}(\tilde{\lambda})\|_2 \le \epsilon .
\]
\end{assumption}

Assumption~\ref{Ass: ass1} states that the HODLR compression error is controlled in spectral norm.
In practice, this can be achieved by choosing a sufficiently small HODLR compression tolerance.

According to Eq. \ref{eq:theta},  the fixed-effect estimators of the exact two-stage LMM with variance ratio $\tilde\lambda$, and the NExt-LMM estimators with variance ratio $\tilde\lambda$ and compression error $\epsilon$ admit the following forms 
\begin{align*}
\tilde{\bm{\theta}}_{\tilde{\lambda}}
&=
\Bigl(W_p^\top \Sigma(\tilde{\lambda})^{-1}W_p\Bigr)^{-1}
W_p^\top \Sigma(\tilde{\lambda})^{-1}y, \\
\tilde{\bm{\theta}}_{\tilde{\lambda},\epsilon}
&=
\Bigl(W_p^\top \tilde{\Sigma}(\tilde{\lambda})^{-1}W_p\Bigr)^{-1}
W_p^\top \tilde{\Sigma}(\tilde{\lambda})^{-1}y.
\end{align*}
The fixed-effect estimators of the two-stage LMM with the true variance ratio $\lambda_0$ is:
\[
\tilde{\bm{\theta}}_{\lambda_0}
=
\Bigl(W_p^\top \Sigma(\lambda_0)^{-1}W_p\Bigr)^{-1}
W_p^\top \Sigma(\lambda_0)^{-1}y .
\]

According to Eq. \ref{eq:sigma}, the estimated variance parameters of the two-stage LMM and NExt-LMM with variance ratio $\tilde\lambda$ and compression error $\epsilon$ are 
\begin{align*}
    \tilde{\sigma}_{e, \tilde\lambda}^2 &= \frac{1}{n-c-1} (\textbf{y} - W_p\tilde{\bm{\theta}}_{ \tilde\lambda})^T{\Sigma}( \tilde\lambda)^{-1}(\textbf{y} - W_p\tilde{\bm{\theta}}_{ \tilde\lambda}), \\
     \tilde{\sigma}_{e, \tilde\lambda, \epsilon}^2 &= \frac{1}{n-c-1} (\textbf{y} - W_p\tilde{\bm{\theta}}_{\tilde\lambda, \epsilon})^T{\tilde\Sigma}(\tilde\lambda)^{-1}(\textbf{y} - W_p\tilde{\bm{\theta}}_{\tilde\lambda, \epsilon}), 
\end{align*}
and the estimated variance parameters of the two-stage LMM with the true variance ratio $\lambda_0$ is defined by:
\begin{align*}
    \tilde{\sigma}_{e, \lambda_0}^2 &= \frac{1}{n-c-1} (\textbf{y} - W_p\tilde{\bm{\theta}}_{\lambda_0})^T{\Sigma}(\lambda_0)^{-1}(\textbf{y} - W_p\tilde{\bm{\theta}}_{\lambda_0}). 
\end{align*}

{
\begin{definition}
\label{def:well_conditioned}
We say that $\Sigma(\tilde{\lambda})$ is uniformly well-conditioned if there exist constants
$0<m\le M<\infty$ such that, for all admissible $\tilde{\lambda}$,
\[
m \le \mu_{\min}\!\bigl(\Sigma(\tilde{\lambda})\bigr)
\le
\mu_{\max}\!\bigl(\Sigma(\tilde{\lambda})\bigr)
\le M .
\]
where $\mu_{\min}(\cdot)$ and $\mu_{\max}(\cdot)$ denote the smallest and largest
eigenvalues of a symmetric matrix, respectively.
\end{definition}}

\begin{definition}
\label{def:preliminary_criterion}
Let $\mathcal I\subset(0,\infty)$ denote the admissible parameter space for the
variance ratio, and let $\mathcal L_n(\lambda)$ be the preliminary null-model
criterion used to estimate the variance ratio, {the definition of $Q$}. The preliminary estimator is defined as
\[
\tilde{\lambda}
\in
\arg\min_{\lambda\in\mathcal I}
\mathcal L_n(\lambda).
\]
The population counterpart of this criterion is denoted by
\[
\mathcal L(\lambda)
=
\lim_{n\to\infty}\mathcal L_n(\lambda),
\]
whenever the limit exists.
\end{definition}

\begin{proposition}
\label{prop:epsilon}
Under Assumption~\ref{Ass: ass1}, suppose that
$\Sigma(\tilde{\lambda})$ is uniformly well-conditioned.
Then, as $\epsilon\to 0$, we have
\begin{align*}
\bigl\|\tilde{\bm{\theta}}_{\tilde{\lambda},\epsilon}
-
\tilde{\bm{\theta}}_{\tilde{\lambda}}\bigr\|_2
&\to 0,
\\
\bigl|
\tilde{\sigma}^2_{e,\tilde{\lambda},\epsilon}
-
\tilde{\sigma}^2_{e,\tilde{\lambda}}
\bigr|
&\to 0.
\end{align*}
In particular, the HODLR-based NExt-LMM estimators converge to the exact shared-$\lambda$
two-stage estimators as $\epsilon\to 0$, with linear perturbation bounds in $\epsilon$. Biologically, this means that as the approximation error vanishes,
NExt-LMM preserves the same marker-effect estimates and residual variation as the
exact two-stage LMM.
\end{proposition}

\begin{assumption}
\label{Ass: ass2}
The admissible parameter space $\mathcal I$ is compact, and the true
variance ratio $\lambda_0$ lies in the interior of $\mathcal I$. The preliminary
null-model criterion $\mathcal L_n(\lambda)$ satisfies the following conditions:
\begin{enumerate}
    \item The empirical criterion converges uniformly to its population counterpart:
    \[
    \sup_{\lambda\in\mathcal I}
    \left|
    \mathcal L_n(\lambda)-\mathcal L(\lambda)
    \right|
    \to 0~ \text{as}~n\to \infty .
    \]

    \item The population criterion $\mathcal L(\lambda)$ is continuous on
    $\mathcal I$ and is uniquely minimized at $\lambda_0$:
    \[
    \mathcal L(\lambda_0)
    <
    \mathcal L(\lambda),
    \qquad
    \text{for all } \lambda\in\mathcal I,\ \lambda\neq\lambda_0 .
    \]

    \item The null-model covariance matrices are uniformly well-conditioned over
    $\mathcal I$, namely, for every $\lambda\in\mathcal I$, all eigenvalues of
    $\Sigma(\lambda)$ are bounded below and above by positive constants independent of
    $\lambda$ and $n$.
\end{enumerate}
\end{assumption}
If the conditions of Assumption \ref{Ass: ass2} are satisfied, then the preliminary variance-ratio estimator satisfies as $n\to \infty$, 
\[
|\tilde{\lambda}-\lambda_0| \to 0 .
\]
For the perturbation analysis below, we denote the realized preliminary estimation error by
\[
\delta := |\tilde{\lambda}-\lambda_0|.
\]
Thus, Assumption~\ref{Ass: ass2} implies $\delta\to 0$ as $n\to\infty$.

Assumption~\ref{Ass: ass2} quantifies the plug-in error induced by using the preliminary estimate
$\tilde{\lambda}$ in place of the true variance ratio $\lambda_0$ when constructing the covariance matrix
for the subsequent two-stage analysis. The PCGC regression provides one example of such a preliminary variance-ratio estimation
procedure, which converges to a population criterion uniquely minimized at the true variance
component. Therefore, the induced variance-ratio estimate satisfies
Assumption~\ref{Ass: ass2}.

\begin{proposition}
\label{prop:delta}
Under Assumption~\ref{Ass: ass2}, as $n \to \infty$, we have
\begin{align*}
\bigl\|\tilde{\bm{\theta}}_{\tilde{\lambda}}
-
\tilde{\bm{\theta}}_{\lambda_0}\bigr\|_2
&\to 0,\\
\bigl|
\tilde{\sigma}^{2}_{e,\tilde{\lambda}}
-
\tilde{\sigma}^{2}_{e,\lambda_0}
\bigr|
&\to 0 .
\end{align*}
\end{proposition}

Proposition~\ref{prop:delta} shows that replacing the true variance ratio
$\lambda_0$ by its preliminary estimate $\tilde{\lambda}$ induces an asymptotically
negligible perturbation in the exact two-stage LMM fixed-effect and residual-variance
estimators.

\begin{proposition}
\label{prop:total}
Under Assumptions~\ref{Ass: ass1} and~\ref{Ass: ass2}, we have
\begin{align*}
\bigl\|
\tilde{\bm{\theta}}_{\tilde{\lambda},\epsilon}
-
\tilde{\bm{\theta}}_{\lambda_0}
\bigr\|_2
&\to 0,
\\
\bigl|
\tilde{\sigma}^{2}_{e,\tilde{\lambda},\epsilon}
-
\tilde{\sigma}^{2}_{e,\lambda_0}
\bigr|
&\to 0 ,
\end{align*}
as 
$\epsilon\to 0$ and $n \to \infty$.
\end{proposition}

Proposition~\ref{prop:total} shows that the combined effects of the preliminary
variance-ratio estimation error and the HODLR covariance approximation error vanish
as $\epsilon\to0$ and $n\to \infty$. Consequently, the NExt-LMM fixed-effect and
residual-variance estimators converge to their exact two-stage LMM counterparts
based on the true shared variance ratio $\lambda_0$. Thus, asymptotically, the computational approximation introduced by NExt-LMM does
not alter the estimated marker effects or residual variation relative to the exact
shared-$\lambda_0$ two-stage LMM. 
}

\subsection{Rank--accuracy tradeoff of the HODLR approximation}
\label{sec:rank_accuracy_tradeoff}

{
The HODLR rank parameter $k$ plays a central role in the computational behavior
of NExt-LMM. A larger $k$ improves the accuracy of the covariance approximation
but increases the storage and arithmetic cost of HODLR construction,
factorization, and linear solves. Conversely, a smaller $k$ leads to faster
computation at the expense of a larger approximation error. Thus, $k$ should be
viewed as an accuracy--efficiency tuning parameter rather than as a purely
algorithmic constant.}
{
This tradeoff is particularly relevant for GSMs.
In this section, $n$ denotes the number of individuals, so that
$K_n\in\mathbb R^{n\times n}$ is the GSM and
\[
\Sigma_n = I_n+\tilde{\lambda}K_n,
\qquad
\tilde{\lambda}\ge 0,
\]
is the covariance matrix to be approximated. GSMs often reflect population
structure, relatedness, and other shared genetic components among individuals \citep{price2006principal,yang2011gcta}.
Such latent structure can lead to off-diagonal numerical compressibility after
hierarchical partitioning, which is precisely the structure exploited by HODLR methods. }
{
We now examine the influence of the parameters $k$ and $n$ on the computational error. Let a dyadic HODLR partition with
\[
L_n:=\lceil \log_2 n\rceil
\]
levels be applied to $\Sigma_n$. At level $\ell$, the index set
$\{1,\ldots,n\}$ is partitioned into $2^\ell$ contiguous index clusters. Each
cluster is further split into two child clusters, producing off-diagonal sibling
blocks. We denote these off-diagonal blocks by
\[
\{B_{\ell,r}\}_{r=1}^{R_\ell},
\]
where $r$ indexes the off-diagonal blocks at level $\ell$ and $R_\ell$ is the
number of such blocks. Each block $B_{\ell,r}$ is approximated by a rank-$k$
matrix $B_{\ell,r}^{(k)}$.
}
{
\begin{definition}
\label{def:hodlr_tail}
For a fixed HODLR rank $k$, define the worst blockwise truncation tail as
\[
e_{n,k}
:=
\max_{0\le \ell\le L_n-1}\ \max_r\
\mu_{k+1}(B_{\ell,r}),
\]
where $\mu_{k+1}(B_{\ell,r})$ is the $(k+1)$-th singular value of the
off-diagonal block $B_{\ell,r}$. By the Eckart--Young--Mirsky theorem \citep{stewart1998perturbation, dax2014low, breiding2018geometry},
$\mu_{k+1}(B_{\ell,r})$ is the smallest possible spectral-norm error among all
rank-$k$ approximations to $B_{\ell,r}$. Thus, $e_{n,k}$ measures the worst
best-achievable rank-$k$ approximation error across all off-diagonal blocks in
the HODLR partition.
\end{definition}
}
{
\begin{assumption}
\label{Ass: hodlr_compression}
There exists a constant $c_{\mathrm{cmp}}\ge 1$ such that for every level
$\ell$ and block index $r$, the computed rank-$k$ approximation satisfies
\[
\|B_{\ell,r}-B_{\ell,r}^{(k)}\|_2
\le
c_{\mathrm{cmp}}\,\mu_{k+1}(B_{\ell,r}).
\]
\end{assumption}
This assumption does not assert that GSMs are
universally low-rank; it only states that, for a given off-diagonal block, the
computed rank-$k$ approximation is comparable to the best rank-$k$ approximation
of that block \citep{martinsson2016compressing}.}

{
\begin{proposition}
\label{prop:global_hodlr_error}
Under Assumption~\ref{Ass: hodlr_compression}, let
$\widetilde{\Sigma}_{n,k}$ be the HODLR approximation obtained by replacing each
off-diagonal block $B_{\ell,r}$ with its rank-$k$ approximation
$B_{\ell,r}^{(k)}$. Then
\[
\|\Sigma_n-\widetilde{\Sigma}_{n,k}\|_2
\le
c_{\mathrm{cmp}}L_n e_{n,k}
=
O\!\left(\log n\cdot e_{n,k}\right).
\]
\end{proposition}
Proposition~\ref{prop:global_hodlr_error} shows that the approximation accuracy
is governed not by the nominal rank $k$ alone, but by the numerical truncation
tail $e_{n,k}$ of the observed covariance structure. For a target spectral-norm
tolerance $\epsilon>0$, define the tolerance-dependent HODLR numerical rank
\[
k_\epsilon(n)
:=
\min\left\{
k:
c_{\mathrm{cmp}}L_n e_{n,k}\le \epsilon
\right\}.
\]
Then \(k_\epsilon(n)\) is the rank required to approximate \(\Sigma_n\) within
tolerance \(\epsilon\) under the HODLR partition. Accordingly, the HODLR complexity
should be further interpreted as
$O\!\left(k_\epsilon(n)^2 n\log^2 n\right)$. This formulation clarifies the computational gain of NExt-LMM depends on the off-diagonal numerical compressibility of the observed GSM-induced covariance matrix.
}

\section{Numerical experiments}
\label{sec:sim}

We utilized \textit{gG2P} software~\citep{tang2019g2p}, a specialized genome-wide association study simulation tool, to generate synthetic genetic variation data from random populations for our computational experiments. The simulation pipeline implemented bi-allelic model through random allocation according to specified genetic parameters. {In addition, the SNP-specific minor allele frequencies were sampled from the empirical MAF spectrum of a real GWAS dataset, namely EGAD00000000028 from the European Genome-phenome Archive (EGA; \url{https://ega-archive.org/datasets/EGAD00000000028}), which was also used in the Section \ref{sec:real}. Conditional on the sampled allele frequency \(p_j\), genotypes were generated as \(G_{ij}\mid p_j \sim \mathrm{Binomial}(2,p_j)\). This revision yielded a substantially more realistic simulation design while avoiding unnecessary tuning parameters.}

 The simulation framework was configured with the following key parameters summarized in Table \ref{tab: g2p}. The notations of this section were the same as described in Section \ref{sec:med}.

\begin{table*}[h!t]
  \begin{center}
    \caption{Key parameters for the simulation of genetic variation by \textit{gG2P}. }
    \label{tab: g2p}
    \begin{tabular}{cccccccc}
    \toprule[1.5pt]
       \multicolumn{2}{c}{\textbf{Parameter}}   &\multicolumn{4}{|c}{\textbf{Symbol	Value/Range}}   &\multicolumn{2}{|c}{\textbf{ Interpretation}}
       \\
        \toprule[1.5pt]
        \multicolumn{2}{c|}{Sample size}   & \multicolumn{4}{|c}{$n\in \{500,1000,5000,10000\}$}    & \multicolumn{2}{|c}{Number of  individuals}
       \\
        \multicolumn{2}{c|}{Marker size}   & \multicolumn{4}{|c}{$P = 60,000$ (fixed)} &  \multicolumn{2}{|c}{Number of genotypes}
       \\
        \multicolumn{2}{c|}{Chromosomes}   & \multicolumn{4}{|c}{ $Ch = 1$ (fixed)}    & \multicolumn{2}{|c}{Simulated genetic unit}
       \\
        \multicolumn{2}{c|}{Heterozygosity}   & \multicolumn{4}{|c}{$ H_e = 0.5
$ (default)}   & \multicolumn{2}{|c}{Heterozygous probability}
       \\
        \bottomrule[1.5pt]
    \end{tabular}
  \end{center}
\end{table*}

\subsection{Simulation 1}
\label{sub:3.1}
In the first simulation study, we systematically evaluated the computational speed and numerical accuracy of HODLR inversion for the covariance matrices $\Sigma = \lambda K + I_n$ under varying genetic architectures. Here $K$ denoted the GSM. \AAA{We parameterized $\lambda = h^2/(1 - h^2)$, so that $\lambda \geq 0$. Since $K $ is positive semi-definite, the covariance matrix $\Sigma = I_n + \lambda K$ is positive definite.} 
We simulated the genotype matrices $X$ containing $P=60,000$ independent SNPs for sample sizes $n \in \{500, 1000, 5000, 10000\}$. $100$ genetic matrices $X$ were generated for each sample size. In addition, we selected four levels of heritability $h^2 \in \{0.1, 0.3, 0.5, 0.8\}$ to represent weak to strong heritability. The total number of generated covariance matrices $\Sigma$ was $1,600$.

\begin{figure*}
  \centering
  \includegraphics[width=\linewidth]{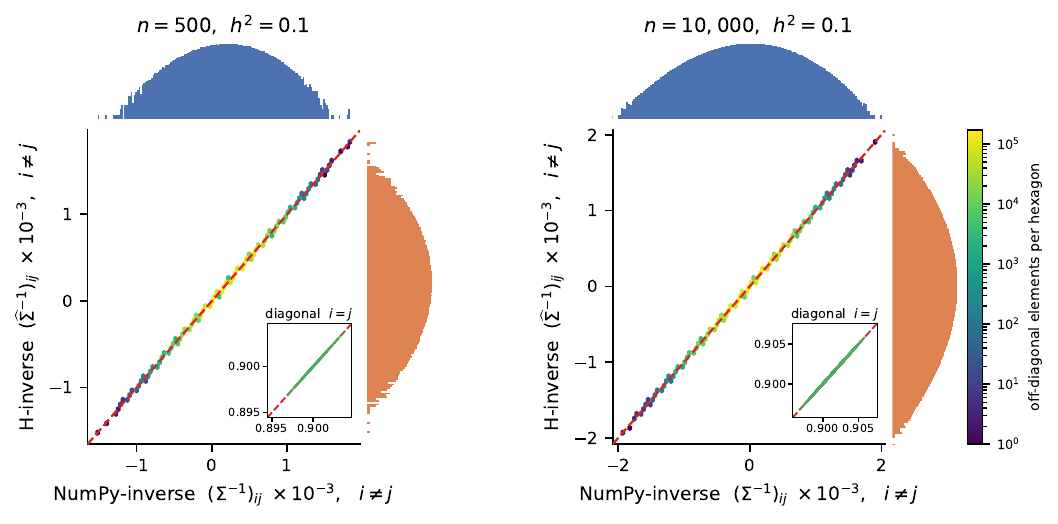}
  \caption{\AAA{Element-wise comparison between the dense \textit{NumPy}-inverse
  and the H-inverse ($h^2=0.1$, $P=60{,}000$, $\epsilon=10^{-3}$, $s=3$). Each
  point is one entry $(\Sigma^{-1})_{ij}$: the horizontal axis gives the value
  from the dense \textit{NumPy}-inverse and the vertical axis the corresponding
  value from the H-inverse. Main panels show off-diagonal entries ($i\neq j$) and
  insets the diagonal entries on separate scales. Hexagonal bins are coloured by
  the number of entries on a logarithmic scale; the top and right histograms give
  the marginal distributions of the horizontal and vertical axes; the red dashed
  line is $y=x$. The quantitative agreement is as follows. At $n=500$ the
  relative Frobenius error of the inverse is $1.5\times10^{-6}$ over all entries
  and $1.5\times10^{-4}$ over the off-diagonal entries alone, the inverse
  residual $\|\Sigma\widehat{\Sigma}^{-1}-I_n\|_F/\sqrt{n}$ is
  $1.5\times10^{-6}$, and the Pearson correlation and least-squares slope
  between the two sets of values are both $1.000000$ to six decimal places; at
  $n=10{,}000$ the corresponding values are $1.6\times10^{-6}$,
  $3.9\times10^{-5}$, $1.5\times10^{-6}$, $1.000000$ and $1.000000$. The
  off-diagonal-only figure is the larger of the two because the diagonal entries
  dominate $\|\Sigma^{-1}\|_F$. The full grid over $n$ and $h^2$ is tabulated in
  Appendix~H of the Supplementary Material.}}
  \label{fig:0.1}
\end{figure*}

We first compared the H-inverse (the HODLR approximations) with the NumPy-inverse (the inversion with the Python package \text{NumPy} \citep{harris2020array}, regarded as the ground truth) in terms of accuracy. 
\AAA{Figure \ref{fig:0.1} displays the element-wise comparison of the inverse
covariance matrices $\Sigma^{-1}$ provided by these two methods, with
$h^2 = 0.1$. Each plotted point is one entry $(\Sigma^{-1})_{ij}$: the horizontal
axis gives the value obtained from the \textit{NumPy}-inverse and the vertical
axis the corresponding value obtained from the H-inverse. Because the diagonal
entries are of order $10^{-1}$--$10^{0}$ whereas the off-diagonal entries are of
order $10^{-3}$, displaying both on a common linear scale makes the off-diagonal
structure invisible; The main panels therefore show the off-diagonal entries ($i\neq j$) and the insets show the diagonal entries on separate scales. We employed hexagonal binning to mitigate over-plotting of points with close coordinates, with the colour of each bin giving the number of entries it contains on a logarithmic scale. The reference line $y=x$ in red denotes perfect agreement
between methods, and the marginal histograms on the top and right axes display
the frequency distributions of the \textit{NumPy}-inverse and H-inverse values,
respectively. Across sample sizes the hexagons align tightly with the reference
line, and the two marginal distributions agree in both central tendency and
dispersion, indicating element-wise concordance of the two methods. Only
$n=500$ and $n=10,000$ are shown here; the remaining sample sizes behave
identically and are summarized quantitatively below.}

\AAA{Scatter plots alone do not quantify the agreement, so we further measured
the relative Frobenius error of the inverse
$\|\widehat{\Sigma}^{-1}-\Sigma^{-1}\|_F/\|\Sigma^{-1}\|_F$, where
$\widehat{\Sigma}^{-1}$ denotes the H-inverse, together with the inverse residual
$\|\Sigma\widehat{\Sigma}^{-1}-I_n\|_F/\sqrt{n}$, which measures how well the
approximate inverse satisfies the defining property of an inverse and is
normalized by $\sqrt{n}=\|I_n\|_F$ so as to be comparable across sample sizes.
Over the full grid $n \in \{500, 1000, 5000, 10000\}$ and
$h^2 \in \{0.1, 0.3, 0.5, 0.8\}$ at tolerance $\epsilon=10^{-3}$, both quantities
are at most $1.5\times10^{-5}$ and agree with each other to within a few
percent, so the product $\Sigma\widehat{\Sigma}^{-1}$ is close to the identity to
the same order as $\widehat{\Sigma}^{-1}$ is close to $\Sigma^{-1}$. The error
increases mildly with $h^2$, consistent with the increasing condition number of
$\Sigma$, while the scale-free relative error is nearly constant in $n$. The mean
absolute element-wise error (MAE) instead \emph{decreases} with $n$, because the
individual entries of $\Sigma^{-1}$ themselves become smaller as $n$ grows; since
the relative error stays nearly constant, no systematic error accumulation across
increasingly deep HODLR partitions is observed. Finally, we assessed the numerical accuracy of the recursive block inversion used to
compute $\tilde{\Sigma}^{-1}$. Specifically, the inverse residual
$\|\tilde{\Sigma}\tilde{\Sigma}^{-1}-I_n\|_F/\sqrt{n}\approx 1.4\times 10^{-14}$
remains at floating-point level across the tested values of $\epsilon$.
Hence, the discrepancy with respect to the original covariance $\Sigma$
is driven by the low-rank compression $\Sigma\approx\tilde{\Sigma}$,
rather than by numerical errors in inverting the compressed matrix.
The complete parameter grid is provided in Appendix H of the Supplementary Material.}

\begin{figure}
    \centering
    \includegraphics[width=0.8\linewidth]{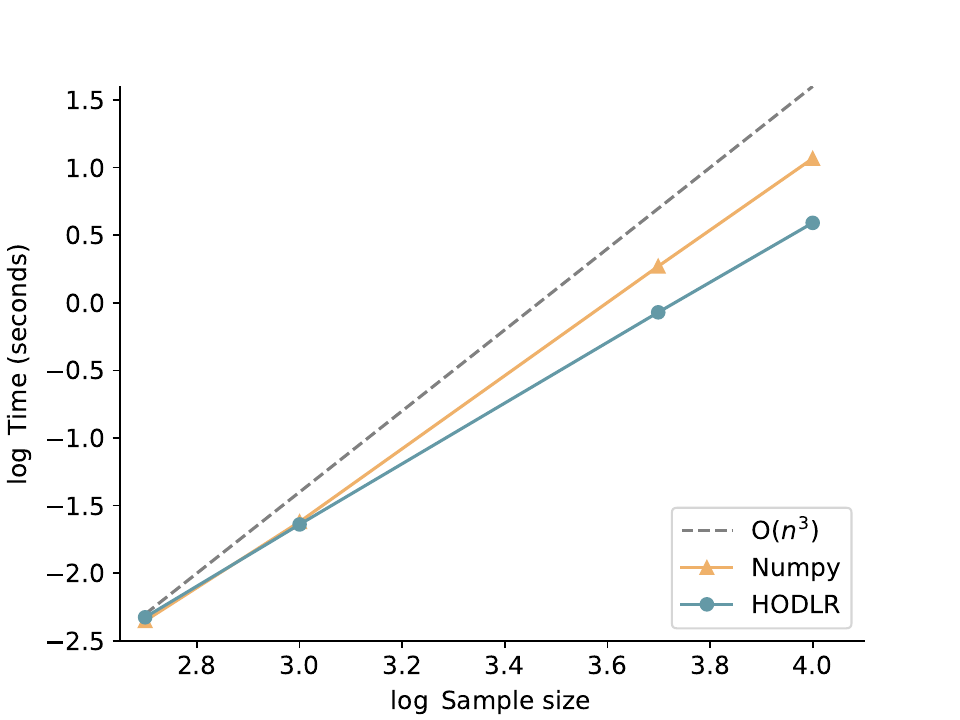}
    \caption{\AAA{Empirical inversion times of the H-inverse and the dense \textit{NumPy} inverse on the unstructured benchmark. Both axes are shown on logarithmic scales. The black dotted line indicates cubic scaling for reference.}}
    \label{fig:complexity}
\end{figure}

We also investigated the acceleration of computing speed achieved by our method.
\AAA{Figure~\ref{fig:complexity} compares the inversion times of the HODLR
procedure and the dense \textit{NumPy} inverse on this unstructured benchmark.
The theoretical direct inverse follows the expected cubic growth, whereas the
H-inverse curve grows more slowly over the tested range; the \textit{NumPy}
inverse sits between the two, since this widely-used package leverages highly
optimized routines and C-accelerated array operations to achieve an empirical
cost below its $\mathcal{O}(n^3)$ asymptotics. At $n=10^4$ the H-inverse is about
$5\times$ faster than the \textit{NumPy} inverse in this benchmark. This
comparison illustrates the potential benefit of hierarchical compression, but it
does not establish near-linear scaling of the hybrid implementation. As shown in
Sections~\ref{sec:rank_emp} and~\ref{sec:sens}, the realized computational
benefit depends on the retained ranks and may require tolerances that are too
coarse for calibrated association inference.}

\subsection{Simulation 2}
\label{sec:sim2}

In this simulation study, we compared NExt-LMM to some state-of-the-art methods in genetic association studies, including the FaST-LMM, fastGWA, and BOLT-LMM.
\AAA{We generate independent genotype matrices $X$ (and the corresponding genetic similarity matrices) according to the replication scheme of Simulation~1}, with sample sizes $n\in \{500,1000,5000,10000\}$. For each $X$, \AAA{the single phenotype vector $\mathbf{y}$} was simulated according to the classical GWAS simulation \citep{yang2011gcta} displayed in the following Equation 
\begin{equation}
   \label{con: sim}
    {\mathbf{y}} = W{\bm{u}} + {\bm{\epsilon}},
\end{equation}
where $W$ denotes the column-standardized matrix of causal variants. The corresponding effect vector $\bm{u}$ follows a sparse mixture of a normal distribution and a Dirac distribution as ${\bm{u}} \sim \pi_1\delta(\textbf{0})+\pi_2\text{MVN}(0,I_P)$, 
with $\pi_1 = 0.95$ and $\pi_2 = 0.05$. The setting implies only about $5\%$ were significant (non-zeros) among total $P=60,000$ SNPs. \AAA{Besides, the residual effect was generated as
\[
\bm{\epsilon}\sim
\text{MVN}\!\left(\mathbf{0},
\frac{\widehat{\operatorname{Var}}(W\bm{u})}{\lambda} I_n\right),
\]
where $\widehat{\operatorname{Var}}(W\bm{u})$ is the empirical variance of the
genetic component and $\lambda = h^2/(1-h^2)$. This construction yields the
target realized heritability $h^2$.} 

 In particular, assuming a low-heritability regime  in simulations is biologically well motivated for many complex traits, including neuropsychiatric phenotypes and brain connectivity measures. First, complex traits are extremely polygenic: thousands of common variants of very small effect contribute to risk, which naturally yields modest SNP-based heritability. This architecture is supported by the omnigenic model and by LD-dependent analyses showing that negative (purifying) selection constrains effect sizes of common variants, keeping $h^2$ low \citep{boyle2017expanded,gazal2017linkage, o2019extreme}. Second, empirical estimates confirm that many relevant traits lie in the $0.05$--$0.15$ range. For example, large-scale GWAS report SNP heritability around $0.09$ for major depressive disorder and around $0.12$ for neuroticism, consistent with highly polygenic, environmentally modulated phenotypes \citep{wray2018genome, nagel2018meta}. Imaging genetics studies likewise find modest SNP heritability for many brain imaging-derived phenotypes, with functional connectivity and other dynamic measures typically lower than structural traits, reflecting strong environmental and state dependence \citep{elliott2018genome, grasby2020genetic}. Taken together, these biological and empirical lines of evidence make $h^2 = 0.1$ a realistic and conservative setting for benchmarking methods under the genetic architectures commonly encountered in practice. Thus in the following simulation, we focus on the low-heritability regime where $h^2=0.1$.

{
We examine the performance of LMM methods by conducting 100 experimental replicates as a function of sample sizes $n$. The results are summarized in Table~\ref{tab:runtime_auc_smalln},  which reports the total runtime and AUC of NExt-LMM and three competing methods (\ie~fastGWA, BOLT-LMM, and FaST-LMM) on simulated datasets with $P=60,000$ SNPs across varying sample sizes. \AAA{In the GWAS analysis, AUC is computed at the marker level as follows. A marker is labeled \emph{positive} if its simulated marker-specific effect is nonzero, and \emph{negative} otherwise; the ranking score is $-\log_{10}(p_j)$ for marker $j$. We compute AUC using the exact Mann--Whitney rank formulation \citep{hanley1982meaning}, with ties handled by averaging ranks. Since the labels correspond to the direct causal markers, a linkage-disequilibrium (LD) proxy of a causal variant is still labeled \emph{negative} even if it exhibits a genuine marginal signal; therefore AUC and marker-level FDR quantify how well the causal markers themselves are ranked, rather than discovery performance.} Overall, all methods showed improved AUC as the sample size increased, indicating enhanced association detection accuracy with larger cohorts.}

{
In terms of computational efficiency, fastGWA achieved the shortest runtime across all sample sizes. However, this computational advantage was accompanied by consistently lower AUC values than NExt-LMM. Our method, NExt-LMM, exhibited favorable scalability, with runtime increasing from 0.04 minutes at $n=500$ to \AAA{$0.72$} minutes at $n=10,000$. Although NExt-LMM was slightly slower than fastGWA, it was substantially faster than BOLT-LMM and FaST-LMM at various sample sizes. For example, at $n=10,000$, NExt-LMM required \AAA{0.72} minutes, compared with \AAA{1.45} minutes for BOLT-LMM and \AAA{2.31} minutes for FaST-LMM.}

{
For estimation accuracy, \AAA{NExt-LMM achieved the highest AUC in all four
sample-size settings, although the margins over BOLT-LMM and FaST-LMM were
small. Specifically, NExt-LMM obtained AUC values of $0.62$, $0.65$, $0.71$ and
$0.74$ as $n$ increased from $500$ to $10,000$, against $0.61$--$0.73$ for
BOLT-LMM, $0.60$--$0.71$ for FaST-LMM and $0.59$--$0.70$ for fastGWA. At
$n=5,000$ NExt-LMM exceeded FaST-LMM by $0.03$ while requiring approximately one
third of its runtime.} \AAA{These AUC values correspond to the low-heritability
regime $h^2=0.1$ on which this section focuses; the substantially higher values
attainable at larger heritability are reported in
Table~\ref{tab:runtime_auc_h2}.}}

{
    These results demonstrate that NExt-LMM provides a strong balance between computational efficiency and association detection performance. While fastGWA is the fastest method, its accuracy is consistently lower than that of NExt-LMM. Conversely, BOLT-LMM and FaST-LMM achieve competitive AUC values but incur substantially higher computational costs. Therefore, NExt-LMM offers an advantageous trade-off, delivering high AUC with markedly reduced runtime relative to the more computationally intensive LMM-based alternatives.
}

{In addition, we examined more extreme heritability settings where the heritability ranges in $[0.05, 0.1, 0.5, 0.9, 0.95]$. The results are shown in Table \ref{tab:runtime_auc_h2}, in which the fastGWA is the most efficient method, followed by NExt-LMM, whereas BOLT-LMM is slower and FaST-LMM is the slowest. In terms of estimation accuracy, NExt-LMM outperforms the other competitors, where fastGWA shows the lowest accuracy. These results further support the conclusion that NExt-LMM achieves a favorable balance between computational efficiency and inferential accuracy across a broad range of heritability regimes. }

\begin{table*}[!t]
\centering
\caption{Comparison of total runtime and AUC for NExt-LMM, fastGWA, BOLT-LMM, and FaST-LMM on simulated data with $P=60,000$ SNPs. Runtime is reported in minutes.
\AAA{All entries correspond to the $h^2=0.1$ configuration, the causal fraction
$\pi=5\%$ and the Wald association statistic, with each metric computed per
phenotype and then averaged over replicates. Runtime covers the complete
end-to-end analysis, that is the construction of $K$, the first-stage
variance-ratio estimation, the covariance factorization and the genome-wide
scan. Because fastGWA, BOLT-LMM and FaST-LMM are all compiled C/C++ programs, the
NExt-LMM runtimes reported here were likewise obtained from a C implementation so
that the comparison is not confounded by low-level implementation differences;
the stage-by-stage timings in Appendix~G of the Supplementary Material are
measured in Python/NumPy instead, in order to isolate the effect of $\epsilon$
and $s$, and are consequently larger. The two sets of runtimes are therefore not
directly comparable.}}
\label{tab:runtime_auc_smalln}
\setlength{\tabcolsep}{6pt}
\renewcommand{\arraystretch}{1.1}
\begin{tabular}{lcccccccc}
\toprule
& \multicolumn{2}{c}{NExt-LMM}
& \multicolumn{2}{c}{fastGWA}
& \multicolumn{2}{c}{BOLT-LMM}
& \multicolumn{2}{c}{FaST-LMM} \\
\cmidrule(lr){2-3} \cmidrule(lr){4-5} \cmidrule(lr){6-7} \cmidrule(lr){8-9}
$n$
& Total (min) & AUC
& Total (min) & AUC
& Total (min) & AUC
& Total (min) & AUC \\
\midrule
500    & 0.04 & \AAA{0.62} & 0.02 & \AAA{0.59} & 0.08 & \AAA{0.61} & 0.13 & \AAA{0.60} \\
1,000  & 0.09 & \AAA{0.65} & 0.03 & \AAA{0.62} & 0.15 & \AAA{0.64} & 0.19 & \AAA{0.63} \\
5,000  & 0.33 & \AAA{0.71} & 0.10 & \AAA{0.67} & 0.64 & \AAA{0.70} & 1.01 & \AAA{0.68} \\
10,000 & \AAA{0.72} & \AAA{0.74} & 0.15 & \AAA{0.70} & \AAA{1.45} & \AAA{0.73} & \AAA{2.31} & \AAA{0.71} \\
\bottomrule
\end{tabular}
\end{table*}

\begin{table*}[!t]
\centering
\caption{Estimated comparison of total runtime and AUC for NExt-LMM, fastGWA, BOLT-LMM, and FaST-LMM on simulated data with fixed $n=10{,}000$ and $P=60{,}000$ SNPs under varying heritability levels. Runtime is reported in minutes.
\AAA{As in Table~\ref{tab:runtime_auc_smalln}, the causal fraction is $\pi=5\%$,
the association statistic is the Wald statistic, and the runtime covers the
complete end-to-end analysis measured from the C implementation of NExt-LMM so
as to match the compiled competitors.}}
\label{tab:runtime_auc_h2}
\setlength{\tabcolsep}{6pt}
\renewcommand{\arraystretch}{1.1}
\begin{tabular}{lcccccccc}
\toprule
& \multicolumn{2}{c}{NExt-LMM}
& \multicolumn{2}{c}{fastGWA}
& \multicolumn{2}{c}{BOLT-LMM}
& \multicolumn{2}{c}{FaST-LMM} \\
\cmidrule(lr){2-3} \cmidrule(lr){4-5} \cmidrule(lr){6-7} \cmidrule(lr){8-9}
$h^2$
& Total (min) & AUC
& Total (min) & AUC
& Total (min) & AUC
& Total (min) & AUC \\
\midrule
0.05 & 0.71 & 0.65 & 0.14 & 0.61 & 1.43 & 0.64 & 2.28 & 0.62 \\
0.10 & 0.72 & 0.74 & 0.15 & 0.70 & 1.45 & 0.73 & 2.31 & 0.71 \\
0.50 & 0.73 & 0.93 & 0.15 & 0.89 & 1.47 & 0.92 & 2.33 & 0.90 \\
0.90 & 0.76 & 0.980 & 0.16 & 0.952 & 1.52 & 0.977 & 2.39 & 0.969 \\
0.95 & 0.77 & 0.986 & 0.16 & 0.958 & 1.55 & 0.982 & 2.41 & 0.974 \\
\bottomrule
\end{tabular}
\end{table*}

{Moreover, we compared NExt-LMM against GEMMA in a high-dimensional \(n\) regime. Specifically, we considered \(n=60{,}000\) and \(100{,}000\) with \(P= 10 \times  n\). These experiments are designed to evaluate whether the approximation used by NExt-LMM preserves inferential accuracy while providing substantial computational gains in the regime that motivates approximate methods. \AAA{Table~\ref{tab:runtime_auc_gemma} compares complete end-to-end runtimes with GEMMA on this unstructured large-$n$ benchmark. NExt-LMM closely matches GEMMA in AUC and has a markedly smaller total runtime under this benchmark and implementation setting. This end-to-end comparison should not be interpreted as a direct measurement of the HODLR solve-stage gain alone: it also reflects the shared-variance two-stage formulation, the respective software implementations, and the complete timing pipeline.} We have also compared the runtime as a function of number of SNPs for a fixed sample size. The result is shown in \AAA{Appendix~C} of the Supplementary Material.}

\begin{table}[htbp]
\centering
\caption{Comparison of total runtime and AUC for NExt-LMM and GEMMA on simulated data with $P=10 \times n$ SNPs. Runtime is reported in hours.
\AAA{The timing convention is that of Table~\ref{tab:runtime_auc_smalln}: the
complete end-to-end analysis, with the NExt-LMM runtimes obtained from the C
implementation so as to match GEMMA.}}
\label{tab:runtime_auc_gemma}
\setlength{\tabcolsep}{5pt}
\renewcommand{\arraystretch}{1.1}
\begin{tabular}{lccccc}
\toprule
& & \multicolumn{2}{c}{NExt-LMM} & \multicolumn{2}{c}{GEMMA} \\
\cmidrule(lr){3-4} \cmidrule(lr){5-6}
$n$ & SNPs ($P$) & Total (h) & AUC & Total (h) & AUC \\
\midrule
60,000  & 600,000   & 1.09 & 0.947 & 71.5  & 0.957 \\
100,000 & 1,000,000 & 3.15 & 0.964 & 330.8 & 0.967 \\
\bottomrule
\end{tabular}
\end{table}

\AAA{\subsection{Simulation 3: structural complexity and the numerical rank of $K$}
\label{sec:rank_emp}

Simulations~1 and~2 use unstructured genotypes, which is the setting in which the
off-diagonal blocks of $K$ are least compressible. Since
Section~\ref{sec:rank_accuracy_tradeoff} shows that the cost of the HODLR
representation is governed by the tolerance-dependent numerical rank
$k_\epsilon(n)$ rather than by a nominal rank, we added a benchmark of $17$
simulated scenarios that vary population differentiation, linkage disequilibrium
(LD) and pedigree relatedness one axis at a time, together with real 1000
Genomes phase~3 genotypes \citep{tgp2015global}, and measured $k_\epsilon$
directly. Three findings are relevant to the method. The rank is driven by the strength of population differentiation rather than by the number of populations; at a fixed $F_{ST}$, adding populations in fact increases the retained rank slightly, consistent with a rank of order $K_{\text{pop}}-1$. LD by itself contributes almost nothing to the numerical rank of $K$, because LD is a correlation between markers whereas $K$ is a covariance between
individuals: an LD-only scenario retains essentially the same off-diagonal ranks as an unstructured one.
The ordering of individuals matters, but only in proportion to the structure that is already present: clustering individuals by ancestry lowers $k_\epsilon$ at the deeper recursion levels by one to two orders of magnitude when differentiation is strong, and has no effect at all when it is absent. Reordering therefore exposes existing compressibility rather than creating it, and does not by itself remove the main computation cost. Accordingly, the HODLR-based speedup is real but conditional: it becomes substantial only when the rank drops a lot. In our experiments, such a large rank drop occurs only when the tolerance is too loose for
calibrated association testing (Section~\ref{sec:sens}).
With the tolerance required for inference, the rank does not drop enough, so the method mainly results in an accuracy--storage--runtime trade-off rather than a clear speedup. The details of analysis and results are given in Appendix~D of the
Supplementary Material.}

\AAA{\subsection{Null calibration, power, and false discovery rate}
\label{sec:calib}
Ranking accuracy alone does not establish the calibration of association
$p$-values. We therefore evaluate NExt-LMM under a polygenic null model on the
benchmarks of Section~\ref{sec:rank_emp}:
\begin{equation}
\mathbf{y}=\mathbf{g}+\boldsymbol{\epsilon},\qquad
\mathbf{g}\sim\mathcal{N}(\mathbf{0},h^2K),\qquad
\boldsymbol{\epsilon}\sim\mathcal{N}(\mathbf{0},(1-h^2)I_n).
\end{equation}
This construction retains the covariance induced by population structure and
relatedness while rendering all tested markers null. For each marker, we use the
Wald statistic
\begin{equation}
\label{eq:wald}
T_p =
\frac{\tilde{\beta}_p}
{\sqrt{\tilde{\sigma}_e^2
\left[
\left(X_p^\top\tilde{\Sigma}^{-1}X_p\right)^{-1}
\right]_{pp}}},
\end{equation}
where $X_p$ contains the covariates and the tested marker, $\tilde{\beta}_p$ is
the estimated coefficient of the tested marker, and $[\,\cdot\,]_{pp}$ denotes
the corresponding diagonal entry. We refer $T_p$ to a $t$ distribution with
$n-c-1$ degrees of freedom, where $c$ is the number of covariates. The
marker-specific standard error accounts for variation in MAF and LD across
markers.}

  \AAA{In this section we set $h^2=0.5$, whereas
  Sections~\ref{sub:3.1}--\ref{sec:sim2} use $h^2=0.1$. The reason is that the
  null calibration of an approximate solver is governed by the weight $K$ carries
  in $\Sigma$: with $\Sigma=I_n+\lambda K$ and $\lambda=h^2/(1-h^2)$, the setting
  $h^2=0.1$ gives $\lambda=1/9$, so that $\Sigma$ is dominated by the identity and
  a perturbation of $K$ is damped by close to an order of magnitude before it can
  reach the test statistic; $h^2=0.5$ gives $\lambda=1$, so that the genetic and
  residual components carry equal weight and the condition number of $\Sigma$ is
  larger (Section~\ref{sub:3.1}). The higher heritability is therefore the more
  demanding setting for the question at issue, namely whether compressing $K$
  perturbs the null distribution.}

\begin{figure*}
  \centering
  \includegraphics[width=\linewidth]{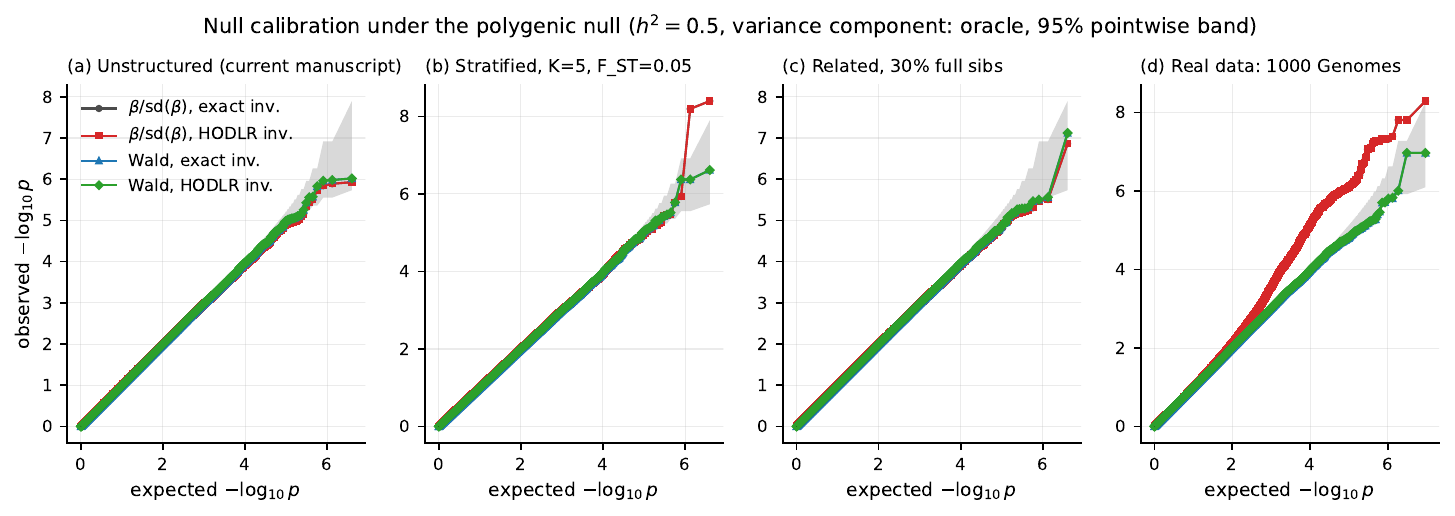}
  \caption{\AAA{QQ plots of null $p$-values under the polygenic null ($h^2=0.5$,
  variance ratio fixed at the simulated value); the grey band is the pointwise
  $95\%$ envelope. In the simulated scenarios the HODLR and exact inverses
  overlap and follow the diagonal. On the real 1000 Genomes genotypes (panel d)
  the statistic of the previous version deviates in the tail, whereas the Wald
  statistic remains calibrated.}}
  \label{fig:r2_qq}
\end{figure*}

\AAA{Figure~\ref{fig:r2_qq} shows that the empirical Type~I error tracks the nominal
level from $\alpha=0.05$ to $10^{-5}$ and that
$\lambda_{\mathrm{GC}}$ remains close to one. At $\epsilon=10^{-3}$, the HODLR
and exact inverses yield nearly identical QQ curves on both simulated and real
genotypes. This agreement is retained when the variance ratio is estimated
separately for each phenotype by REML.}

\AAA{We further evaluate power and realized false discovery rate (FDR) under the
genome-wide threshold $p<5\times10^{-8}$ and the
Benjamini--Hochberg (BH) procedure at $q=0.05$. The experiments span causal
fractions
\[
\pi\in\{0.01\%,\,0.1\%,\,1\%,\,5\%\},
\]
as well as a mixed architecture containing a small number of moderate effects
and many weak effects. Table~\ref{tab:r2_power} reports locus-level power and
FDR for all settings, together with marker-level FDR in the final column. The
two FDR definitions coincide in the simulated scenarios without LD, whereas
their difference on the real-genotype benchmark quantifies discoveries
attributable to LD proxies. Under
fixed total heritability, denser architectures distribute genetic variance over
more causal markers and therefore yield lower marginal power. The realized FDR
is near the nominal level in most moderate and dense settings, but can exceed
$0.05$ for ultra-sparse architectures and for real genotypes with heterogeneous
LD. We therefore report realized FDR for every setting and do not interpret BH
at $q=0.05$ as guaranteeing finite-sample locus-level FDR control.}

\AAA{For AUC and PR-AUC, positive labels correspond to simulated causal markers and
negative labels to all remaining markers, with $-\log_{10}(p_p)$ used as the
ranking score. AUC is computed from its exact Mann--Whitney rank formulation.
Because an LD proxy of a causal marker is labelled negative under this
marker-level definition, AUC quantifies ranking of direct causal markers rather
than LD-aware discovery performance; the latter is additionally assessed using
locus-level power and FDR. Full simulation protocols, replicate aggregation,
null-calibration ranges, FDR definitions, and additional diagnostic results are
provided in Supplementary Appendix~E.}

\begin{table*}[!t]
\centering
\caption{\AAA{Power, realized FDR and ranking accuracy versus the genetic
architecture (Wald statistic, HODLR inverse at $\epsilon=10^{-3}$, variance ratio
fixed at the simulated value, $h^2=0.5$). Power and FDR are at the locus level;
the marker-level FDR, which counts markers in LD with a causal variant as false
discoveries, is in the last column. The two definitions coincide in the simulated
scenarios, which contain no LD. ``mixed'' denotes 10 moderate effects plus $1\%$
weak effects; AUC and PR-AUC use the direct causal-marker labels. Averages are
over 5 genotype matrices $\times$ 100 phenotypes for $n=1{,}000$ and
3 $\times$ 50 otherwise.}}
\label{tab:r2_power}
\setlength{\tabcolsep}{5pt}
\renewcommand{\arraystretch}{1.05}
\AAA{\resizebox{\textwidth}{!}{\begin{tabular}{llcccccccc}
\toprule
Scenario & Architecture & Causal loci & Power @$5\times10^{-8}$ & Power @BH & FDR @BH & AUC & PR-AUC & FDR @BH \\
 & & & (locus) & (locus) & (locus) & & & (marker) \\
\midrule
\multirow{5}{*}{S1c, $n=1{,}000$}
 & $\pi=0.01\%$ & $2$ & $0.843$ & $0.866$ & $0.234$ & $0.977$ & $0.876$ & $0.234$ \\
 & $\pi=0.1\%$ & $20$ & $0.292$ & $0.411$ & $0.062$ & $0.880$ & $0.506$ & $0.062$ \\
 & $\pi=1\%$ & $199$ & $2.9\times10^{-3}$ & $0.018$ & $0.054$ & $0.687$ & $0.110$ & $0.054$ \\
 & $\pi=5\%$ & $997$ & $1\times10^{-5}$ & $1.2\times10^{-4}$ & $0.038$ & $0.563$ & $0.074$ & $0.038$ \\
 & mixed & $209$ & $6\times10^{-3}$ & $0.019$ & $0.040$ & $0.672$ & $0.100$ & $0.040$ \\
\midrule
\multirow{5}{*}{S1c, $n=5{,}000$}
 & $\pi=0.01\%$ & $2$ & $0.903$ & $0.913$ & $0.042$ & $0.991$ & $0.934$ & $0.042$ \\
 & $\pi=0.1\%$ & $20$ & $0.626$ & $0.711$ & $0.019$ & $0.948$ & $0.777$ & $0.019$ \\
 & $\pi=1\%$ & $200$ & $0.118$ & $0.276$ & $0.021$ & $0.834$ & $0.464$ & $0.021$ \\
 & $\pi=5\%$ & $999$ & $2.3\times10^{-3}$ & $0.025$ & $0.020$ & $0.690$ & $0.248$ & $0.020$ \\
 & mixed & $210$ & $0.103$ & $0.245$ & $0.019$ & $0.822$ & $0.434$ & $0.019$ \\
\midrule
\multirow{5}{*}{S6, $n=2{,}504$ (real)}
 & $\pi=0.01\%$ & $5$ & $0.667$ & $0.773$ & $0.537$ & $0.960$ & $0.301$ & $0.977$ \\
 & $\pi=0.1\%$ & $46$ & $0.203$ & $0.419$ & $0.235$ & $0.863$ & $0.098$ & $0.941$ \\
 & $\pi=1\%$ & $463$ & $2.2\times10^{-3}$ & $0.021$ & $0.046$ & $0.656$ & $0.041$ & $0.675$ \\
 & $\pi=5\%$ & $2316$ & $0$ & $3.3\times10^{-4}$ & $8.5\times10^{-3}$ & $0.548$ & $0.065$ & $0.134$ \\
 & mixed & $473$ & $5.9\times10^{-3}$ & $0.028$ & $0.111$ & $0.644$ & $0.034$ & $0.854$ \\
\bottomrule
\end{tabular}
}}
\end{table*}

\AAA{\subsection{Sensitivity to the HODLR parameters}
\label{sec:sens}

Finally we quantify how the compression tolerance $\epsilon$ and the minimum
block size $s$ affect the approximation error, the computational cost and the
resulting inference, analyzing the same genotype matrices and phenotypes under
each setting and under the exact inverse so that all differences are
attributable to the approximation alone. The tolerance behaves as a single
user-facing control: the forward approximation error decreases approximately
linearly with $\epsilon$, the inverse residual stays of the same order over the
whole grid, and solve time and storage decrease monotonically with $\epsilon$
because they are governed by the retained rank. 

However, the impact of $\epsilon$ on the final inference is not the same as its impact on the computation cost. When $\epsilon=10^{-3}$, the approximation is practically inert in our tests: in the simulations,
no marker changes its genome-wide significance status, and on the real genotypes,
only about one marker per phenotype changes. At this setting, $\lambda_{GC}$, the Type~I error, the power,
and the estimated variance components are the same as in the exact-inverse baseline.
The small difference in the variance components follows the expected scaling with the covariance approximation error
(see Proposition~\ref{prop:epsilon}). By contrast, if we use a looser tolerance (larger $\epsilon$), the ranks collapse and the solve step becomes cheaper.
But on the real genotypes, this already makes $\lambda_{GC}$ and the Type~I error larger. Therefore, $\epsilon$ should be selected based on statistical accuracy, not only on computational speed.
We recommend $\epsilon\leq 10^{-3}$, and we further verify this choice using the approximation error
and the inverse residual on the data at hand, because the real GSM is more sensitive to $\epsilon$
than the simulated ones. 

The minimum block size is not a
substantive tuning parameter here: raising $s$ shortens the recursion but leaves
every other quantity, upstream and downstream, unchanged to the displayed
precision, because $s$ acts only on the deepest levels, whose blocks contribute
negligibly to storage and to the dense inversion cost; we use a cache-friendly
$s=64$ as a practical default. The details are given in Appendix~G of the Supplementary Material.}

\section{Application}
\label{sec:real}

{
We applied our proposed method to a real human genotype dataset from the Precocious Coronary Artery Disease Study (PROCARDIS). Coronary artery disease (CAD) remains one of the leading causes of morbidity and mortality worldwide, and understanding its genetic architecture is a central task in cardiovascular genomics. In the analysis, we used the plasma lipoprotein(a) [Lp(a)] concentration as a continuous quantitative trait for association testing, which is a causally implicated risk factor for atherosclerotic cardiovascular diseases. The dataset consisted of $n=15,937$ subjects recruited from four European countries, including the United Kingdom, Italy, Sweden, and Germany, each with $34,399$ SNPs. The complete set of association results was available from the European Genome--Phenome Archive (\url{https://www.ebi.ac.uk/ega/}) under accession number \text{EGAD00000000028}.}

{
One main methodological challenge in this real data analysis is the potential genetic heterogeneity and population stratification induced by the multicenter European recruitment design. Although all subjects were of European origin, samples were collected from multiple countries, and unmodeled ancestry differences may lead to inflated test statistics or spurious associations. Therefore, a robust linear mixed model framework is required to account for relatedness and latent population structure. Following the original study design, country of origin is treated as a population-level covariate, and the GSM is computed from standardized SNP genotypes to model the confounding structure.}

\begin{figure*}
  \centering
  \subfloat[Kinship]{
     \includegraphics[width=2.6in]{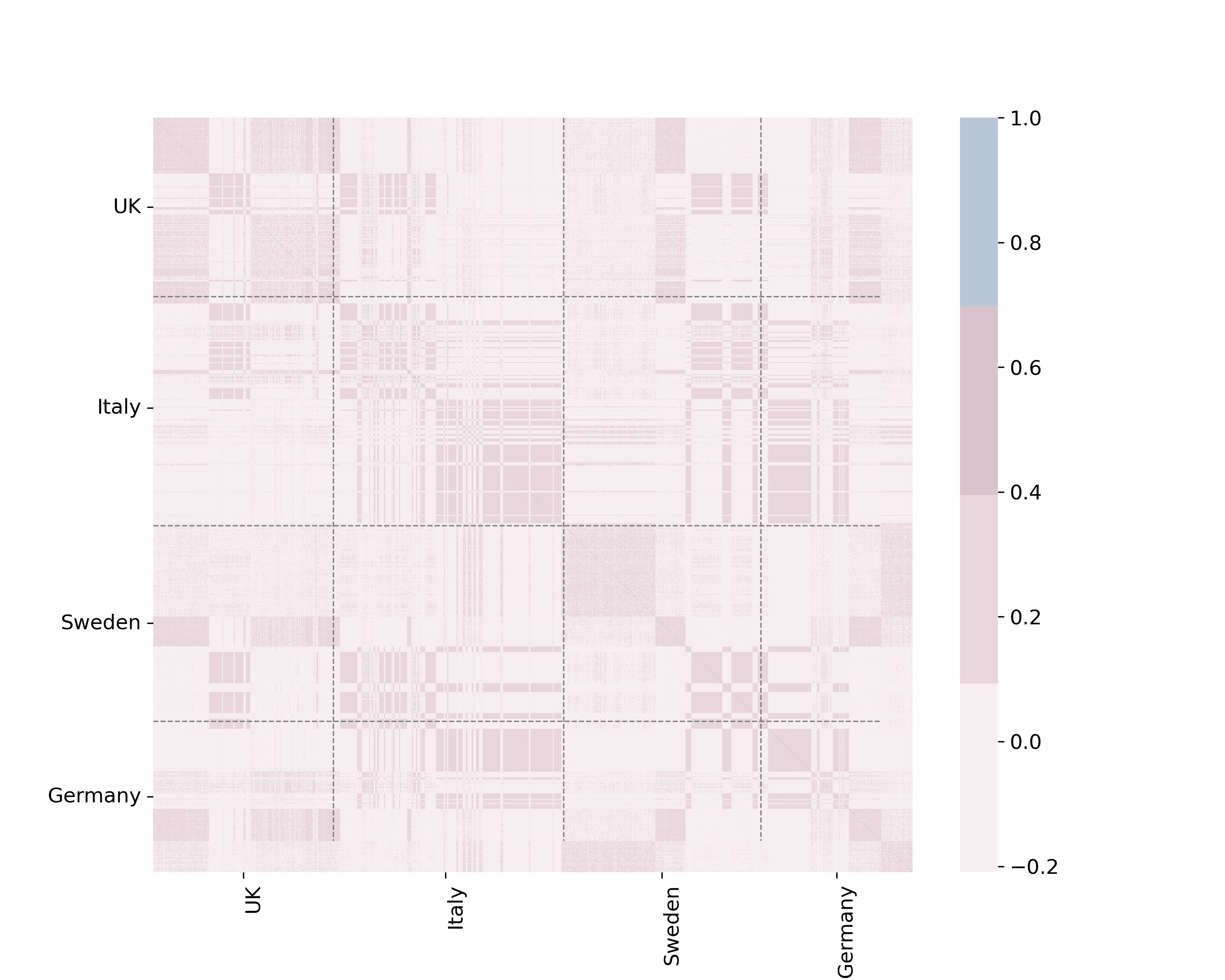}
    \label{subfig:heat}}
  \hfil
    \subfloat[PCA]{
     \includegraphics[width=2.5in]{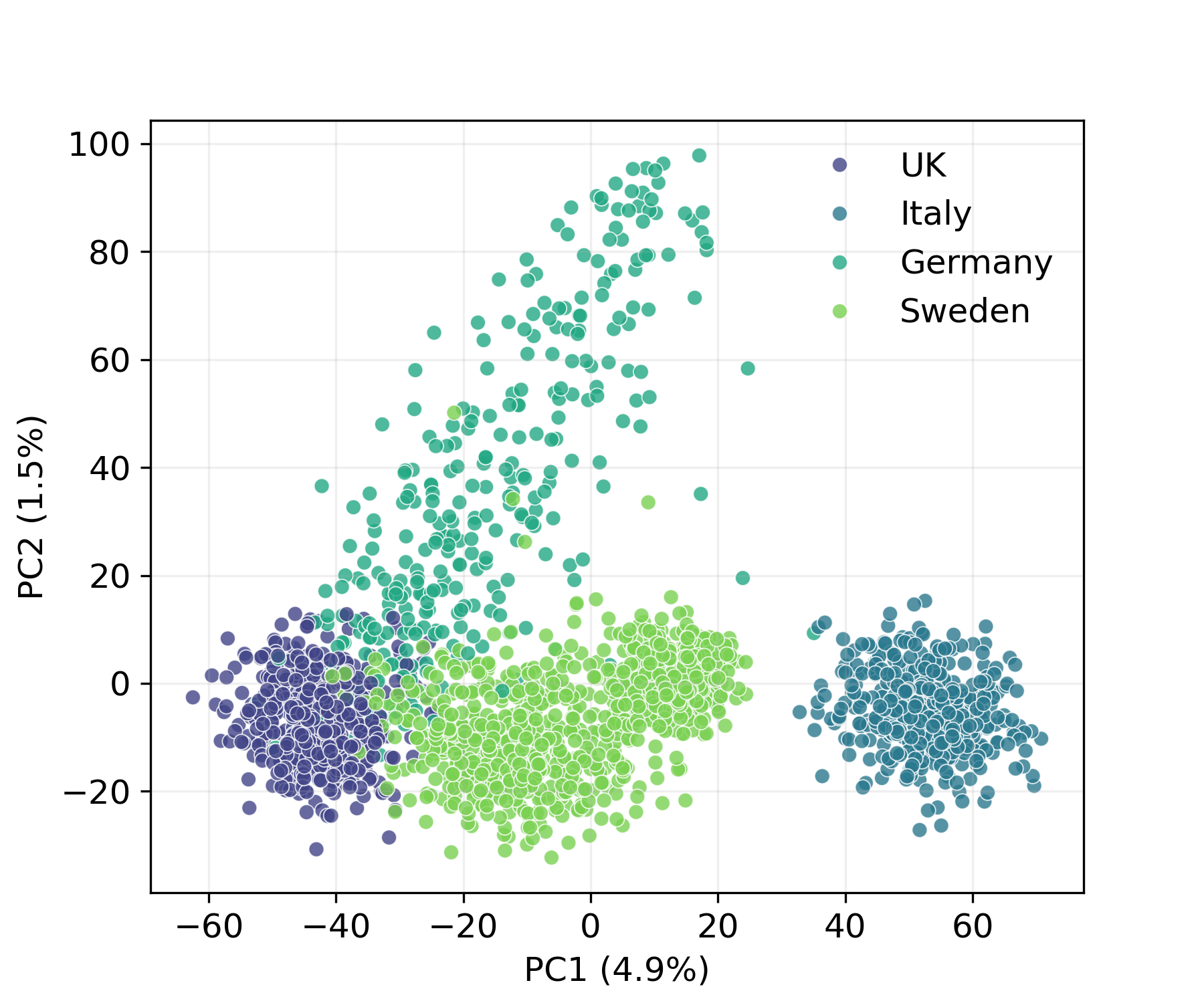}
    \label{subfig:pca}}
  \caption{Population stratification of the  PROCARDIS cohort. (a) shows the heat-map of the kinship matrix, while (b) is the PCA results using the first two most-significant principal components (PCs). Both the sub-figures indicate obvious population stratification of the genotypes. }  
  \label{fig: structure}
\end{figure*}

{
To validate the population structure, we performed the principal component analysis (PCA, Figure \ref{subfig:pca}) on the GSM computed from standardized SNP data and drew the kinship heatmap (Figure \ref{subfig:heat}). These analyses confirm the presence of non-negligible structure among samples, supporting the use of mixed-model-based association testing.}

\begin{figure*}
    \centering
    \includegraphics[width=0.7\linewidth]{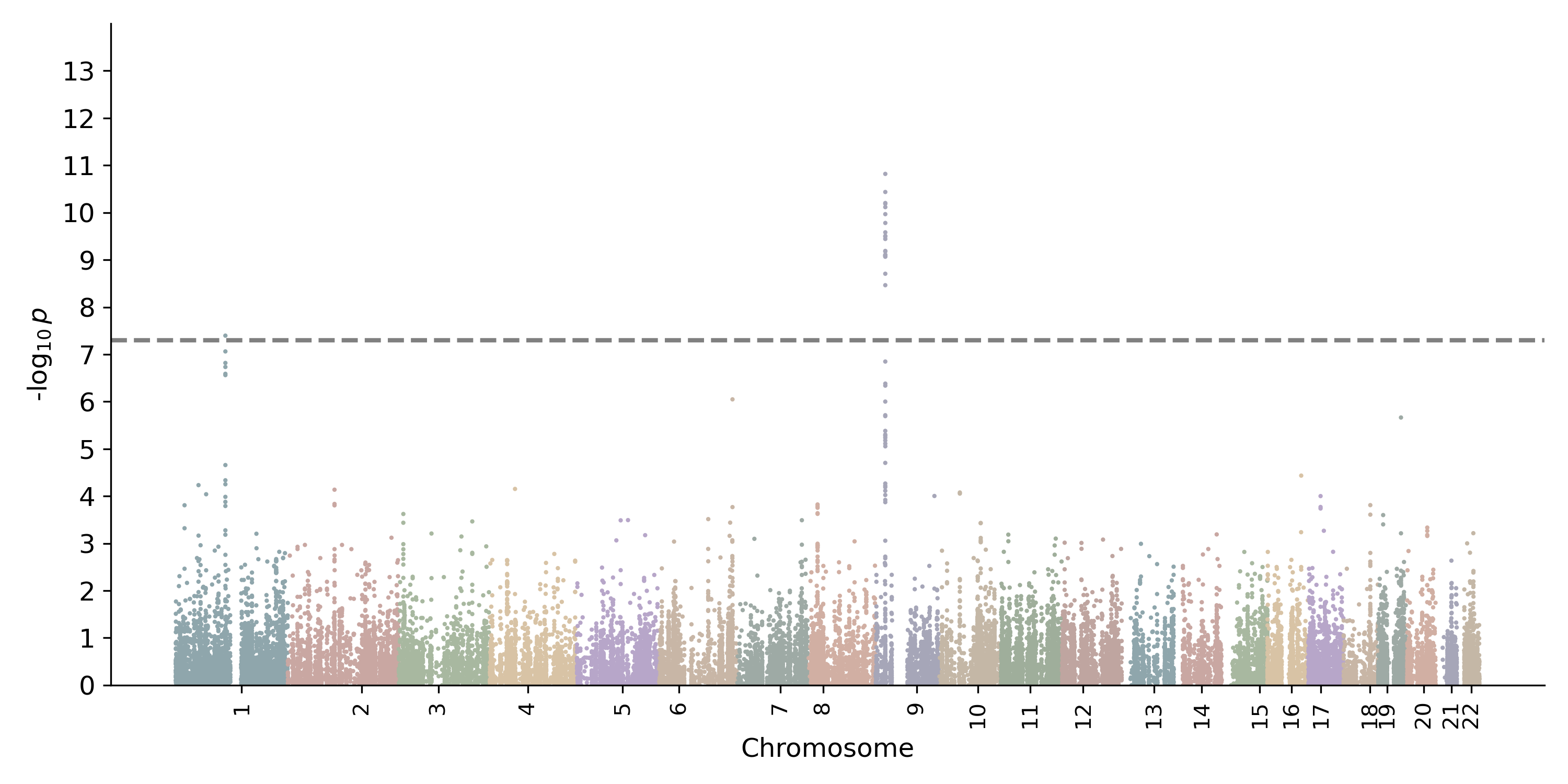}
    
    \caption{Manhattan plot of the testing results on the $34.4$k SNPs by NExt-LMM. $25$ significantly associated loci are detected with the p-value threshold $5 \times 10^{-8}$ (the gray dotted line).}
     \label{fig:next}
\end{figure*}

{
We then applied the NExt-LMM to perform genome-wide association testing while accounting for the confounding structure induced by genetic relatedness and population stratification.  The Manhattan plot (Figure \ref{fig:next}) identified $25$ genome-wide significant loci  $(p < 5 \times 10^{-8})$ across the $34.4$k SNPs by NExt-LMM. We observed high concordance between the NExt-LMM and other methods under comparison, with an average of $22.67$ shared associations across all methods. More precisely, NExt-LMM almost covered the hits detected by the FaST-LMM (which uncovers $21$ hits) with $84\%$ agreement. fastGWA and BOLT-LMM showed $24$ ($96\%$ agreement) and $23$ shared hits ($92\%$ agreement) with NExt-LMM, respectively. Moreover, the NExt-LMM-based result covers all the $23$ significant signals found in \cite{clarke2009genetic}, including rs10455872 and rs3798220 which have been widely considered to be associated with increased Lp(a) levels and elevated CAD risk. These findings demonstrate NExt-LMM' capability to detect genetic signals while controlling for population structure. Furthermore, the computational efficiency (computing time $= 5.4$ minutes for $n = 15,937, P = 34,399$ dataset, at least $2.1 \times$ speed-ups compared to the state of the art methods) enables rapid analysis on moderate or even large scale association studies.
}

\section{Conclusion}
\label{sec:con}

We introduce the NExt-LMM, a computational framework that overcomes critical bottlenecks for the GSM inversion in genome-wide association studies through three contributions. First, we exploit the inherent low-rank structure of GSMs through recursive HODLR approximations. \AAA{For a fully recursive HODLR implementation, and when the retained off-diagonal ranks remain bounded, this reduces the theoretical complexity of the GSM inversion from $\mathcal{O}(n^3)$ to $\mathcal{O}(k_\epsilon^2n\log^2 n)$ while maintaining the user-specified numerical fidelity, where $k_\epsilon$ is the numerical rank retained at the prescribed tolerance. The gain realized in practice depends on the off-diagonal numerical compressibility of the realized GSM and on the chosen individual ordering. Section~\ref{sec:rank_emp} shows that population differentiation can reduce the retained rank at coarse tolerances, whereas LD alone has little effect on the rank of the individual-level GSM. In the present hybrid implementation, the tolerances at which rank reduction yields appreciable solve-stage savings overlap with those at which association calibration deteriorates. Thus, at inference-appropriate tolerances, the observed behavior is primarily an accuracy--storage--runtime trade-off rather than a uniform speedup. We provide two reference implementations; the fully recursive one is designed to attain this complexity under bounded off-diagonal ranks, whereas the numerical experiments reported here use a hybrid implementation that retains a dense terminal block, so that the storage and runtime reported in Section~\ref{sec:sim} should not be interpreted as demonstrating this asymptotic complexity.} Second, we integrate the HODLR-inverse and SNP-shared heritability with the likelihood function. \AAA{On the unstructured baseline benchmarks, the complete end-to-end NExt-LMM pipeline is $1.7$--$3.3\times$ faster than BOLT-LMM and FaST-LMM under the reported implementation and timing convention, although fastGWA remains faster in absolute terms. This comparison reflects the full two-stage analysis pipeline rather than REML estimation alone.} Third, \AAA{we establish perturbation bounds showing that the marker-effect and residual-variance estimators of NExt-LMM converge to those of the exact shared-variance two-stage LMM at a rate linear in the compression tolerance $\epsilon$. Consequently, the inference obtained from the compressed covariance can be brought arbitrarily close to that of the exact two-stage estimator by choosing $\epsilon$. This behavior is also verified numerically.} \AAA{These results motivate further optimization and large-scale evaluation of the fully recursive implementation on biobank-scale cohorts. Such evaluation is necessary before making runtime claims for datasets of the scale of UK Biobank.}


While the NExt-LMM demonstrates the efficacy of HODLR-based approximations for accelerating the inference, there exists several considerations. For the accuracy-complexity tradeoff, the fundamental tension between approximation fidelity and computational efficiency still persists. 
\AAA{A tolerance that is too coarse (\eg~$\epsilon \gg 10^{-3}$) can reduce the retained ranks and empirical solve time, but may inflate Type~I error in association testing. Thus, tolerance selection must be guided by inferential calibration rather than computational cost alone.} \AAA{In addition, because LD can induce proximal contamination in mixed-model association tests, leave-one-chromosome-out (LOCO) analysis is recommended for human-genetic applications when chromosome-specific association testing is required \citep{yang2014advantages}.} \AAA{A practical limitation is that the computational benefit depends on the off-diagonal numerical compressibility of the realized GSM and is therefore not guaranteed for every dataset. In our controlled experiments, population differentiation \citep{price2006principal, astle2009population} was the main driver of reduced rank, whereas LD alone had little effect on the rank of the individual-level GSM. The retained ranks also depend on the ordering of individuals, for which ancestry-aware clustering can be beneficial. More generally, datasets whose GSM blocks remain numerically high rank at the tolerance required for calibrated inference will not obtain substantial computational savings from the present hybrid implementation \citep{ambikasaran2013n, hackbusch1999sparse}.} {Besides, our simulations and theoretical development are based on the standard additive LMM framework. In practice, however, more complex genetic architectures, including dominance, epistasis, and gene--environment interactions, may induce covariance structures that are not fully captured by a single additive genomic relationship matrix. In such settings, the low-rank behavior exploited by NExt-LMM may differ from that observed under purely additive models. Extending the present framework to richer multi-component or interaction-based covariance models is an important direction for future work.}

{We also acknowledge that using a shared heritability estimate across all SNPs is a simplifying assumption. In the current two-stage LMM framework, the variance parameters are estimated once and then used for genome-wide SNP-level testing. This treats SNPs homogeneously with respect to their contribution to heritability and ignores potential differences across functional annotations or genomic regions. In practice, SNPs in coding regions, regulatory elements, conserved regions, or cell-type-specific annotations may have different effect-size distributions or be enriched for heritability. Thus, a common heritability value may be inadequate for capturing annotation-dependent genetic architectures. Extending NExt-LMM to allow SNP-specific or annotation-stratified heritability estimates is an important future direction.}

There are several lines of future research. 
One future research direction is to improve our method by taking advantage of the matrix sparse approximation \citep{schafer2021sparse} and develop the hybrid HODLR-sparse representations with more robust properties. 
Secondly, we would extend the method from single SNP and single trait to fine-mapping \citep{schaid2018genome} to better uncover the hidden patterns of linkage disequilibrium (LD) among SNPs. Finally, the future versions of NExt-LMM will incorporate  computing frameworks with higher performance (\eg~ Intel' OpenMPI libraries) for more efficient applications on the biobank scale data.

\section*{Acknowledgements}
This project is supported by the National Natural Science Foundation of China (No. 12401383, 12101333), the Shanghai Science and Technology Program (No. 21010502500), the startup fund of ShanghaiTech University, and the HPC Platform of ShanghaiTech University. Correspondence send to Drs. Shijia Wang (wangshj1@shanghaitech.edu.cn) and Shufei Ge (geshf@shanghaitech.edu.cn). 

\bibliographystyle{IEEEtran}
\bibliography{IEEEabrv,reference}

\newpage
\begin{IEEEbiographynophoto}{Zhibin Pu} 
Zhibin Pu is a Ph.D. student at ShanghaiTech University. His research interests include Bayesian statistics, statistical machine learning, bioinformatics, and statistical genetics.
\end{IEEEbiographynophoto}%
\vspace{-0.5\baselineskip}
\begin{IEEEbiographynophoto}{Shufei Ge} 
Dr. Shufei Ge received her Ph.D. in statistics from Simon Fraser University in 2020 and joined ShanghaiTech University as an assistant professor that same year. Her research interests lie at the intersection of Bayesian statistics, statistical machine learning, bioinformatics, and statistical genetics. In particular, she focuses on developing scalable statistical methods to address pressing challenges in computational biology, including neuroimaging genetics and complex disease prediction. Her work aims to bridge the gap between advanced statistical theory and practical applications in modern biomedical research.
\end{IEEEbiographynophoto}
\vspace{-0.5\baselineskip}
\begin{IEEEbiographynophoto}{Shijia Wang}
Dr. Shijia Wang is an assistant professor at ShanghaiTech University. He received his PhD in statistics from Simon Fraser University in Canada. His research interests include Bayesian statistics, statistical machine learning, and computational biology.
\end{IEEEbiographynophoto}%
\end{document}